\documentclass[12pt]{article}
\pdfoutput=1

\usepackage{amsmath,amssymb,cases,graphicx,hyperref,multirow,setspace,subfigure}
\usepackage[letterpaper,textheight=8.6in,textwidth=7in]{geometry}

\numberwithin{equation}{section}
\newcommand{\ab}[1]{\left|#1\right|}
\newcommand{\br}[1]{\left[#1\right]}
\newcommand{\cu}[1]{\left\{#1\right\}}
\newcommand{\pa}[1]{\left(#1\right)}

\newcommand{\ed}{\,\mathrm{d}}
\newcommand{\pd}{\,\partial}

\begin{document}

\hfill

\begin{center}
	\vspace{2cm}
	{\LARGE\bf Electromagnetic jets from stars and black holes}\\
	\vspace{2cm}
	Samuel E. Gralla$,\!\!^{\diamondsuit}$ Alexandru Lupsasca,\!$^\spadesuit$ and Maria J. Rodriguez$^\heartsuit$\\
	\vspace{2cm}
	{\it Center for the Fundamental Laws of Nature, Harvard University\\
	Cambridge, Massachusetts 02138, USA}
\end{center}

\vspace{2cm}

\begin{abstract}
	We present analytic force-free solutions modeling rotating stars and black holes immersed in the magnetic field of a thin disk that terminates at an inner radius. The solutions are exact in flat spacetime and approximate in Kerr spacetime. The compact object produces a conical jet whose properties carry information about its nature. For example, the jet from a star is surrounded by a current sheet, while that of a black hole is smooth. We compute an effective resistance in each case and compare to the canonical values used in circuit models of energy extraction. These solutions illustrate all of the basic features of the Blandford-Znajek process for energy extraction and jet formation in a clean setting.
\end{abstract}

\vspace{3cm}

{\let\thefootnote\relax\footnotetext{
	$^{\diamondsuit}$sgralla@physics.harvard.edu
	$\quad^\spadesuit$lupsasca@fas.harvard.edu
	$\quad^\heartsuit$mjrodri@physics.harvard.edu
}}

\pagebreak

\tableofcontents

\pagebreak

\section{Introduction}

The extraction of the spin energy of a rotating compact object by its plasma magnetosphere is a basic physical process that is likely at work in a number of energetic astrophysical phenomena, from pulsar winds and jets to active galactic nuclei (AGN), and possibly including gamma-ray bursts and tidal disruption events. For pulsars, this process is essentially the original unipolar induction of Faraday \cite{Faraday:1832}, in which a conductor rotating in a magnetic field generates an electric field, whose voltage will drive a current if wires are present. The neutron star is the conductor and magnet, while the plasma functions as wires. The study of this process began in the original work of Goldreich and Julian \cite{Goldreich:1969sb} and continues today.

That such a process can also occur for black holes \cite{Blandford:1977ds} is {\it a priori} rather shocking. After all, black holes cannot source a magnetic field, cannot provide charge carriers, and cannot support outflow of local energy. Nevertheless, a number of physical processes conspire to allow black holes to behave as unipolar inductors: an accretion disk can provide the magnetic field, immersing a spinning black hole in such a magnetic field produces electric fields (with nonzero $\vec{E}\cdot\vec{B}$) \cite{Wald:1974np}, these electric fields can pair produce plasma from the vacuum by accelerating stray charges to energies above the pair-creation threshold \cite{Ruderman:1975ju}, and spinning black holes can support outflow of the energy relevant to physical work at infinity \cite{Penrose:1969pc,Penrose:1969,Penrose:1971uk}. Blandford and Znajek (BZ) connected these formerly disparate ideas in their ground-breaking paper \cite{Blandford:1977ds} and demonstrated an astrophysically viable Penrose process---the analog of unipolar induction---with the potential to power AGN jets.

BZ, along with much subsequent work, considered the {\it force-free} approximation for the plasma, in which the matter degrees of freedom do not contribute to the dynamics, leaving a closed, nonlinear evolution system for the electromagnetic field alone \cite{Uchida:1997general,Komissarov:2002my,Gralla:2014yja}. While this is likely a good approximation for much of the magnetosphere (and essentially all of it in the pulsar case), it also has the virtue of being the simplest set of plasma equations that can support the basic mechanism of unipolar induction by stars and black holes in plasma. In particular, while nonlinear, the equations are simple enough to admit {\it analytic solution} in a number of cases. Indeed, in the last few years there has been an explosion of progress on analytic solutions, due in part to new techniques imported from the relativity \cite{Gralla:2014yja,Brennan:2013jla,Brennan:2013ppa,Yang:2014zva,Zhang:2014pla,Zhang:2015aga,Gralla:2015wva,Yang:2015ata} and high-energy theory \cite{Lupsasca:2014pfa,Lupsasca:2014hua,Compere:2015pja} communities. While numerical simulation is essential to gaining enough detail to eventually capture all of the relevant effects in real pulsars and AGN, analytic work is invaluable for elucidating basic mechanisms and efficiently exploring their range of operation.

\begin{figure*}
	\centering
	\subfigure[vertical]{\includegraphics[width=0.24\textwidth]{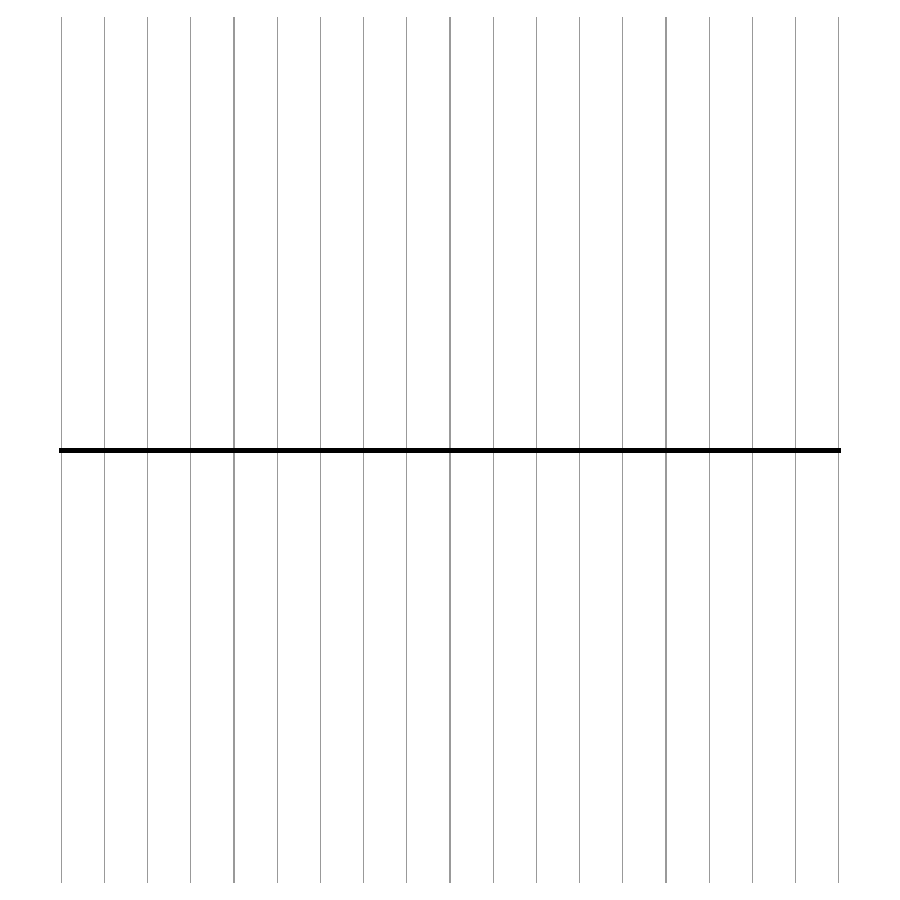}}
	\subfigure[radial]{\includegraphics[width=0.24\textwidth]{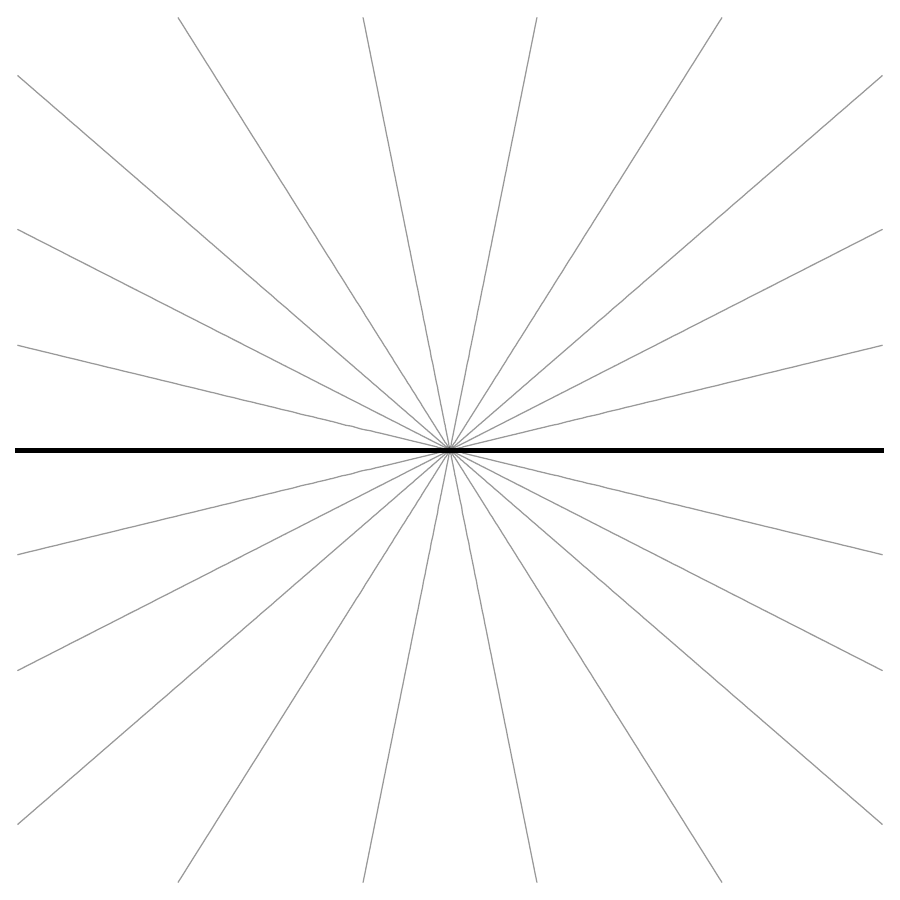}} 
	\subfigure[parabolic]{\includegraphics[width=0.24\textwidth]{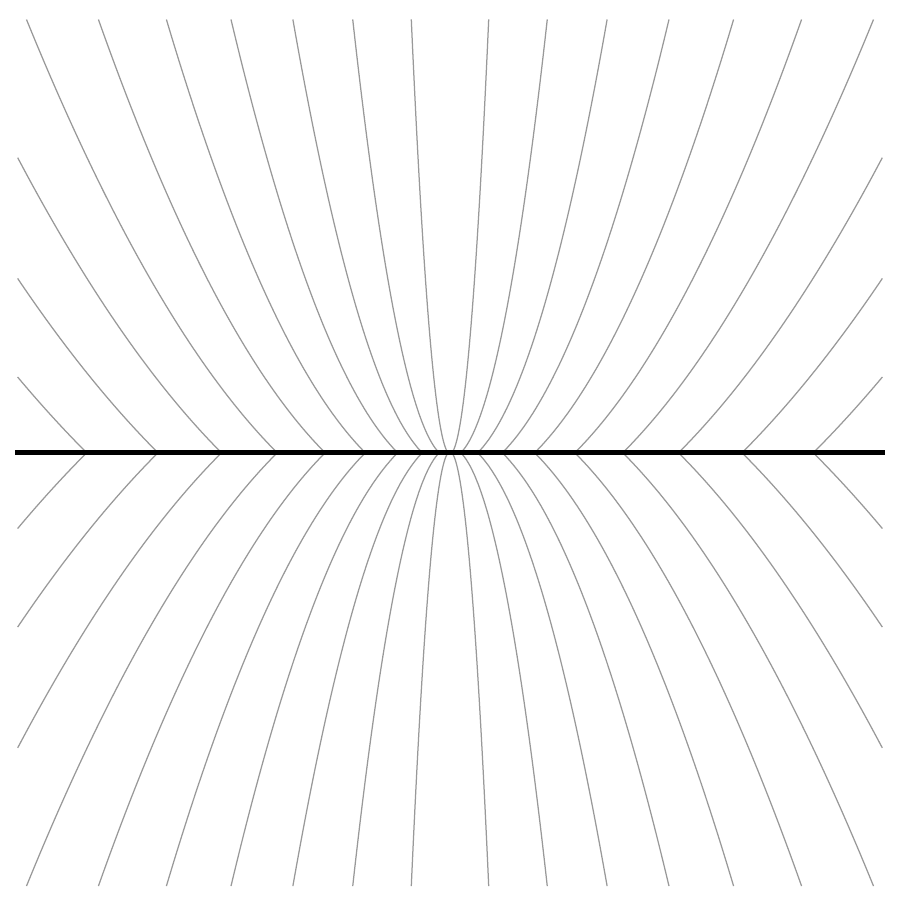}} 
	\subfigure[hyperbolic]{\includegraphics[width=0.24\textwidth]{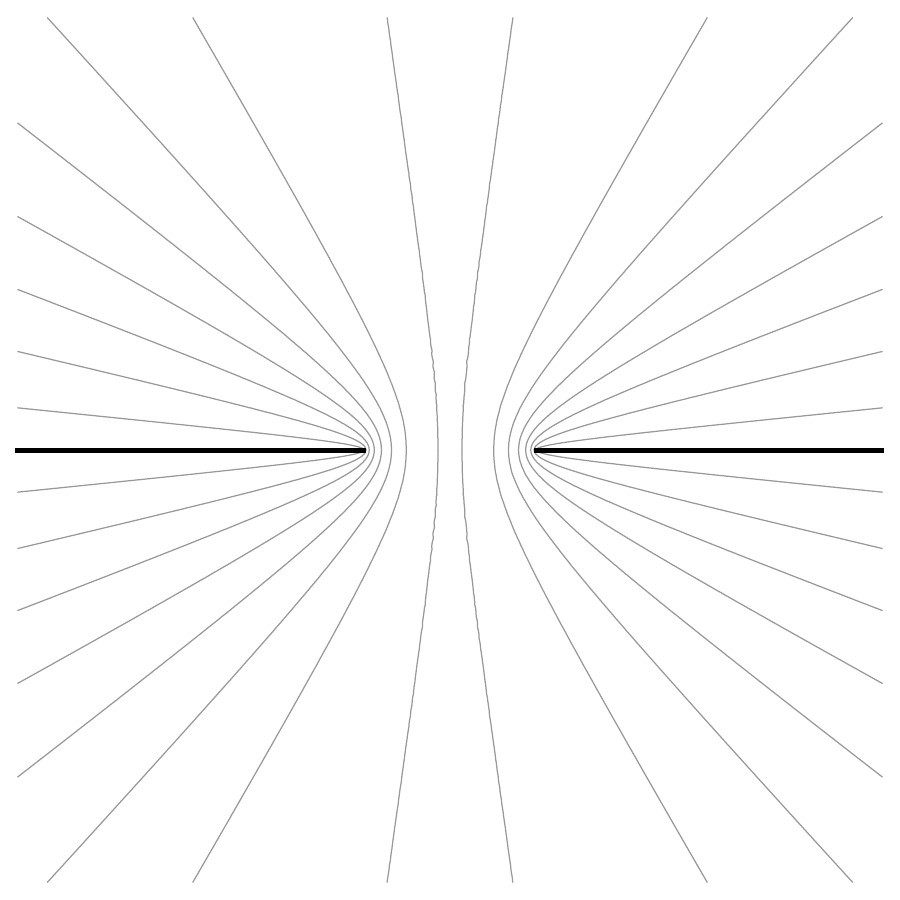}}
	\caption{Poloidal field lines of known families of exact force-free magnetospheres, with current sheets depicted by a thick line. Our new, hyperbolic solution includes the radial and vertical solutions as limits. For the plots we take $\Omega=0$ for simplicity, but the rotating solutions are similar.}
	\label{fig:FieldLineGeometries}
\end{figure*}

In this paper, we continue the effort to push analytic solutions as far as they can be taken. We begin in flat spacetime, which is a decent approximation for neutron stars and, following the method of BZ, a springboard for approximate black hole solutions. For stationary, axisymmetric, energy-extracting, force-free magnetospheres, there are three previously known families of exact solutions, which may be designated by the shapes of their poloidal field lines\footnote{Toroidal refers to the $\phi$ direction, while poloidal refers to the remaining directions: $\rho$ and $z$ in cylindrical coordinates. Poloidal field lines are simply the projection of the magnetic field lines onto the poloidal plane.}: vertical, radial, and parabolic.\footnote{The radial solution was found by Michel \cite{Michel:1973} and the parabolic by Blandford \cite{Blandford:1976}. Vertical solutions appear in the analysis of relativistic jets specialized to full cylindrical symmetry, where the theory becomes an ordinary differential equation that can be directly integrated (e.g., Refs.~\cite{Appl:1993,Lovelace:2012uc}).} In this paper we will add a fourth, ``hyperbolic'' solution, whose basic geometry resembles that of an AGN, with a thin accretion disk sourcing the magnetic field. The four field line geometries are illustrated in Fig.~\ref{fig:FieldLineGeometries}.

The new family of solutions is parametrized by the disk's inner termination radius $b$ as well as a free function $\Omega(\psi)$, corresponding to the angular velocity of each field line $\psi$. (These concepts will be defined precisely below.) We can produce a complete model of unipolar induction, including jet formation, by choosing $\Omega(\psi)$ to equal the angular velocity of a rigidly rotating conducting star (of radius $R$) on field lines that intersect the star and to vanish elsewhere. This generates a conical jet of opening angle $\theta^{\rm star}_{\rm jet}=\arcsin(R/b)$ surrounded by a thin sheet of current flowing in the reverse direction. This model could apply to a neutron star surrounded by a magnetized accretion disk or, as a basic mechanism, to protostellar jets.

\begin{figure*}
	\centering
	\subfigure[star]{\includegraphics[width=0.38\textwidth]{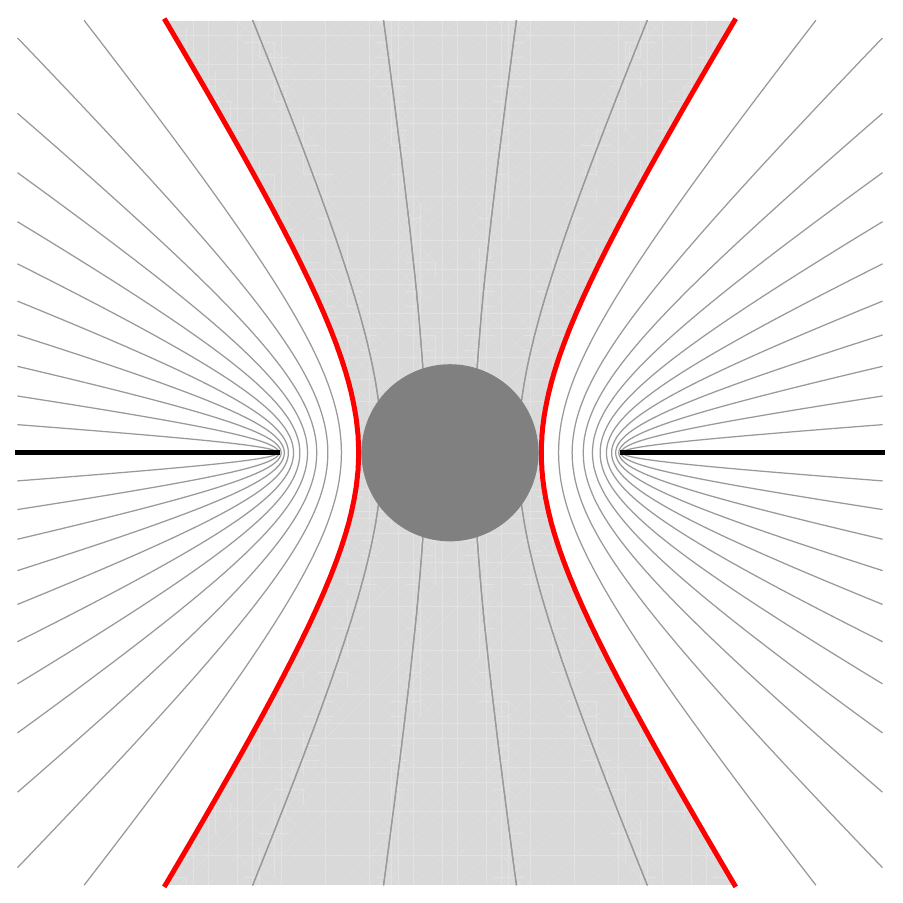}}
	\hspace{1cm}
	\subfigure[black hole]{\includegraphics[width=0.38\textwidth]{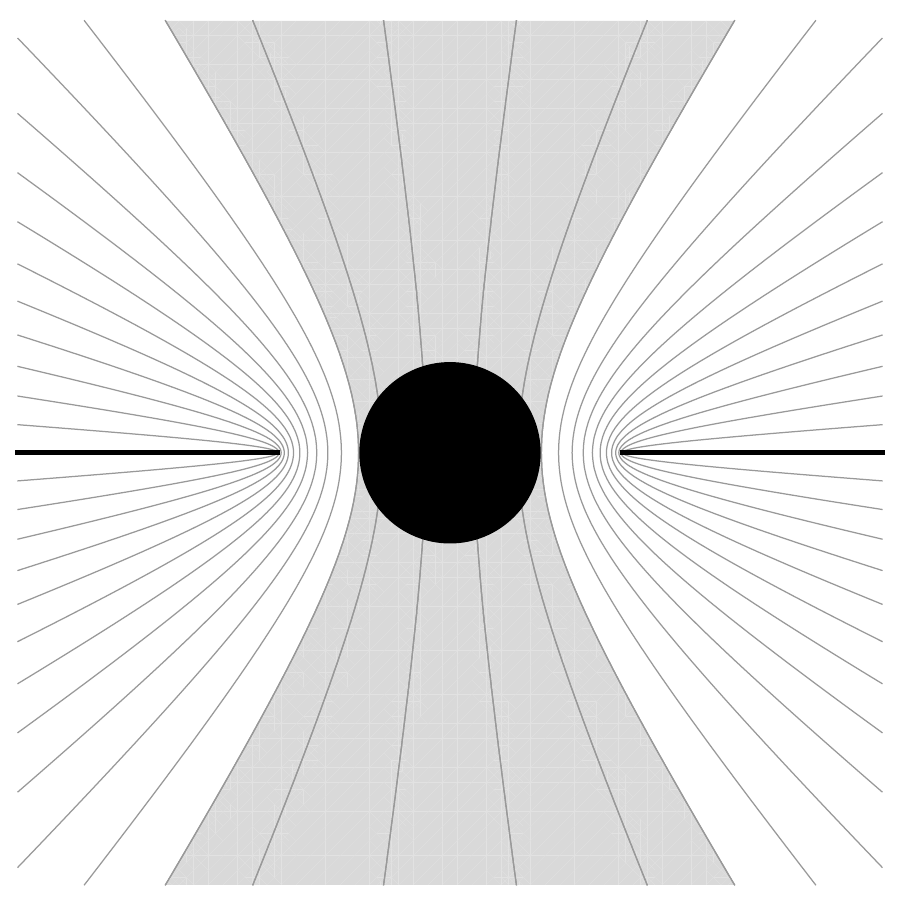}}
	\caption{Jet formation via unipolar induction by rotating compact objects in force-free plasma with magnetic field provided by a thin disk. We provide exact solutions for stars in flat spacetime and perturbative solutions for slowly rotating black holes. The main qualitative difference is that the jet produced by a star is surrounded by a current sheet (shown in red), while the jet produced by a black hole is not.}
	\label{fig:Jets}
\end{figure*}

Following the method of BZ, we can promote this exact flat space solution to an approximate solution for a slowly spinning black hole. Aspects of this procedure were previously carried out by Beskin, Istomin, and Pariev \cite{Beskin:1992,Beskin:2010iba}, who were not in possession of the exact flat solution. Taking the disk to terminate at the innermost stable circular orbit (ISCO) of the black hole, the opening angle is $\approx 20^\circ$, and unlike in the star case, the return current is smoothly distributed in the jet, so that there is no current sheet. One could in principle utilize this qualitative difference to determine the nature of the central object from observations of the jet.\footnote{Note that the jet will only be conical as far away as the force-free approximation holds; beyond that limit, gas confinement mechanisms will likely fix the shape.} Figure~\ref{fig:Jets} illustrates these results.

There has been much discussion over the years of the extent to which black holes behave like conducting bodies. In possession of nontrivial analytic solutions with the same magnetic geometry in each case, we are able to make a direct comparison. In particular, we can test the idea that the black hole has an effective resistance equal to that of free space (equal to $1$ in our units, or approximately $377$ Ohms).\footnote{This notion holds precisely within the membrane paradigm \cite{MacDonald:1982zz,Thorne:1986iy,Penna:2015cga} as a {\it surface} resistivity dictating the behavior of {\it ficticious} electric and magnetic fields (defined by using infinitely accelerated observers with suitable renormalization). The idea of assigning the whole black hole a true resistivity of $1$ has later been applied in models for binaries \cite{McWilliams:2011zi,Lai:2012qe}.} In a limit where the two solutions can be precisely compared, we find that the effective total resistance in the star circuit is equal to $0.5$, while the effective total resistance in the black hole circuit is approximately $0.77$. Thus, the black hole does in a sense provide extra resistance, but it adds a much smaller amount than that of free space.

These solutions also possess pedagogical value. While the main features of the BZ mechanism---an accretion disk sourcing a magnetic field that threads the black hole, producing a jet---are by now well-established, there has not previously existed a complete analytic treatment demonstrating all the features in a single setting. Our treatment also contextualizes the BZ mechanism within the more established physics of unipolar induction. We hope that the calculations presented here will help newcomers understand this beautiful process.

The remainder of the paper is organized as follows. In Sec.~\ref{sec:HyperbolicMagnetosphere}, we derive the new family of exact flat force-free magnetospheres. In Sec.~\ref{sec:Star}, we model unipolar induction by stars with disks, and in Sec.~\ref{sec:BH}, we find approximate solutions in Kerr and model unipolar induction by black holes with disks (i.e., the BZ mechanism). We conclude with a comparison of the star and black hole cases in Sec.~\ref{sec:StarBH}. Our metric signature for flat spacetime is $(-,+,+,+)$ and we work in Heaviside-Lorentz units with $G=c=1$.

\section{Hyperbolic Magnetosphere}
\label{sec:HyperbolicMagnetosphere}

We consider force-free field configurations, i.e., closed 2-forms $F$ ($\nabla_{[a}F_{bc]}=0$) satisfying the force-free condition
\begin{align}
	\label{eq:ForceFreeCondition}
	F_{ab}J^b=0,
\end{align}
where $J^b\equiv\nabla_cF^{bc}$ is the charge-current density four-vector.\footnote{Reviews of force-free electrodynamics, including the machinery used here for stationary, axisymmetric solutions, may be found in Refs.~\cite{Beskin:2010iba} (primarily the $3+1$ approach) and \cite{Gralla:2014yja} (primarily the covariant approach).} This is the statement of vanishing Lorentz four-force density, $\rho\vec{E}+\vec{J}\times\vec{B}=0$ and $\vec{J}\cdot\vec{E}=0$. We restrict our attention to stationary and axisymmetric solutions with nonzero poloidal magnetic field, which may always be written \cite{Gralla:2014yja,Uchida:1997symmetry} as
\begin{align}
	\label{eq:CylindricalAnsatz}
	2\pi F=\ed\psi\wedge\br{\ed\phi-\Omega(\psi)\ed t}+\frac{I(\psi)}{\rho}\ed z\wedge\ed\rho,
\end{align}
where $(t,z,\rho,\phi)$ are cylindrical coordinates and we use the orientation $\ed t\wedge\ed z\wedge\ed\rho\wedge\ed\phi$.\footnote{The electric and magnetic fields are given by $2\pi\vec{E}=-\Omega\vec{\nabla}\psi$ and $2\pi\vec{B}=\rho^{-1}(\vec{\nabla}\psi\times\hat{\phi}+I\hat{\phi})$. Our conventions differ from Ref.~\cite{Gralla:2014yja} (GJ) by $\psi=2\pi\psi_{\rm GJ}$ and from Ref.~\cite{Beskin:2010iba} (B) by $\psi=\psi_{\rm B}/\sqrt{4\pi}$, $I=-\sqrt{4\pi}I_{\rm B}$, $E=E_{\rm B}/\sqrt{4\pi}$, and $B=B_{\rm B}/\sqrt{4\pi}$, where the $\sqrt{4\pi}$ accounts for the change to Gaussian units.\label{Footnote}} Here, $\psi(\rho,z)$ is the magnetic flux function, defined to be the upward\footnote{In this paper, by ``upward'' we mean using a surface that intersects the symmetry axis in the northern hemisphere and has orientation induced by the right-hand rule on the bounding loop.} magnetic flux through the loop of revolution (about the symmetry axis) at the point $(\rho,z)$ in the poloidal plane. The level sets of $\psi$ are the poloidal projections of the magnetic field lines, or poloidal field lines for short; these lines are also equipotential surfaces. The quantity $I$ is the total electric current flowing upward through the $(\rho,z)$ loop. That $I$ is only a function of $\psi$ corresponds to the statement that the poloidal current flows along poloidal magnetic field lines. The toroidal magnetic field is also proportional to $I$ by $B_{\hat{\phi}}=I/(2\pi\rho)$. Finally, $\Omega(\psi)$ represents the angular velocity of each field line $\psi$.\footnote{One physical consequence of this notion is that collisionless particles making up the plasma will be stuck to the rotating line. It is also convenient because field lines that intersect rotating conductors must corotate (see Sec.~\ref{sec:Star}). See (e.g.) Ref.~\cite{Gralla:2014yja} for further discussion of the notion of field line angular velocity.} 

The energy flowing upward through a loop labeled by $\hat{\psi}$ is given by
\begin{align}
	\label{eq:Flux}
	P=-\frac{1}{2\pi}\int_{0}^{\hat{\psi}}I(\psi)\Omega(\psi)\ed\psi.
\end{align}
The differential voltage drop across a field line is $\ed V=\Omega(\psi)\ed\psi$. Thus we have $\ed P\sim I\ed V$ at a differential level, which is reminiscent of the circuit equation $P=IV$. However, here $I$ is the {\it total} current enclosed, while $\ed P$ and $\ed V$ are the {\it local} energy flux and voltage drop rates.

The form of Eq.~\eqref{eq:CylindricalAnsatz} ensures that three of the four components of the force-free condition \eqref{eq:ForceFreeCondition} are satisfied. The remaining component in flat space is the so-called stream equation for $\psi$ \cite{Michel:1973,Scharlemann:1973,Okamoto:1974},
\begin{align}
	\label{eq:StreamEquation}
	\pa{1-\rho^2\Omega^2}\nabla^2\psi-\frac{2}{\rho}\pd_\rho\psi-\rho^2(\nabla\psi)^2\Omega\Omega'+II'=0,
\end{align}
where a prime denotes a derivative with respect to $\psi$. If definite expressions for $I(\psi)$ and $\Omega_F(\psi)$ are chosen, Eq.~\eqref{eq:StreamEquation} becomes a second-order, nonlinear\footnote{Choices of $I$ and $\Omega$ that linearize the equation have various physical difficulties (e.g., Sec.~2.6.2 of Ref.~\cite{Beskin:2010iba}).} partial differential equation for $\psi$. For analytic solutions, this is not a particularly promising approach, since nonlinear partial differential equations are generally difficult to solve. Furthermore, it is hard to see how to choose physically interesting $I(\psi)$ and $\Omega(\psi)$ before determining $\psi$ itself.

An alternative approach consists of introducing an auxiliary variable $u(\rho,z)$ and assuming that $\psi$, $I$, and $\Omega$ are only functions of $u$,
\begin{align}
	\label{eq:Unicorn}
	\psi=\psi(u),\qquad
	I=I(u),\qquad
	\Omega=\Omega(u).
\end{align}
This makes the level sets of $\psi$ the same as those of $u$, so that by picking $u$ one chooses the {\it shape} of the poloidal field lines but not their density. There is no guarantee that a force-free solution corresponding to a particular shape will exist, but at least one can choose physically interesting shapes and hope for the best. This is the method used by Blandford \cite{Blandford:1976} to find the parabolic solution and by Menon and Dermer \cite{Menon:2005mg} to find their null solution in Kerr. The radial and vertical solutions can also be derived straightforwardly this way.

\begin{figure}
	\centering
	\includegraphics[scale=0.6]{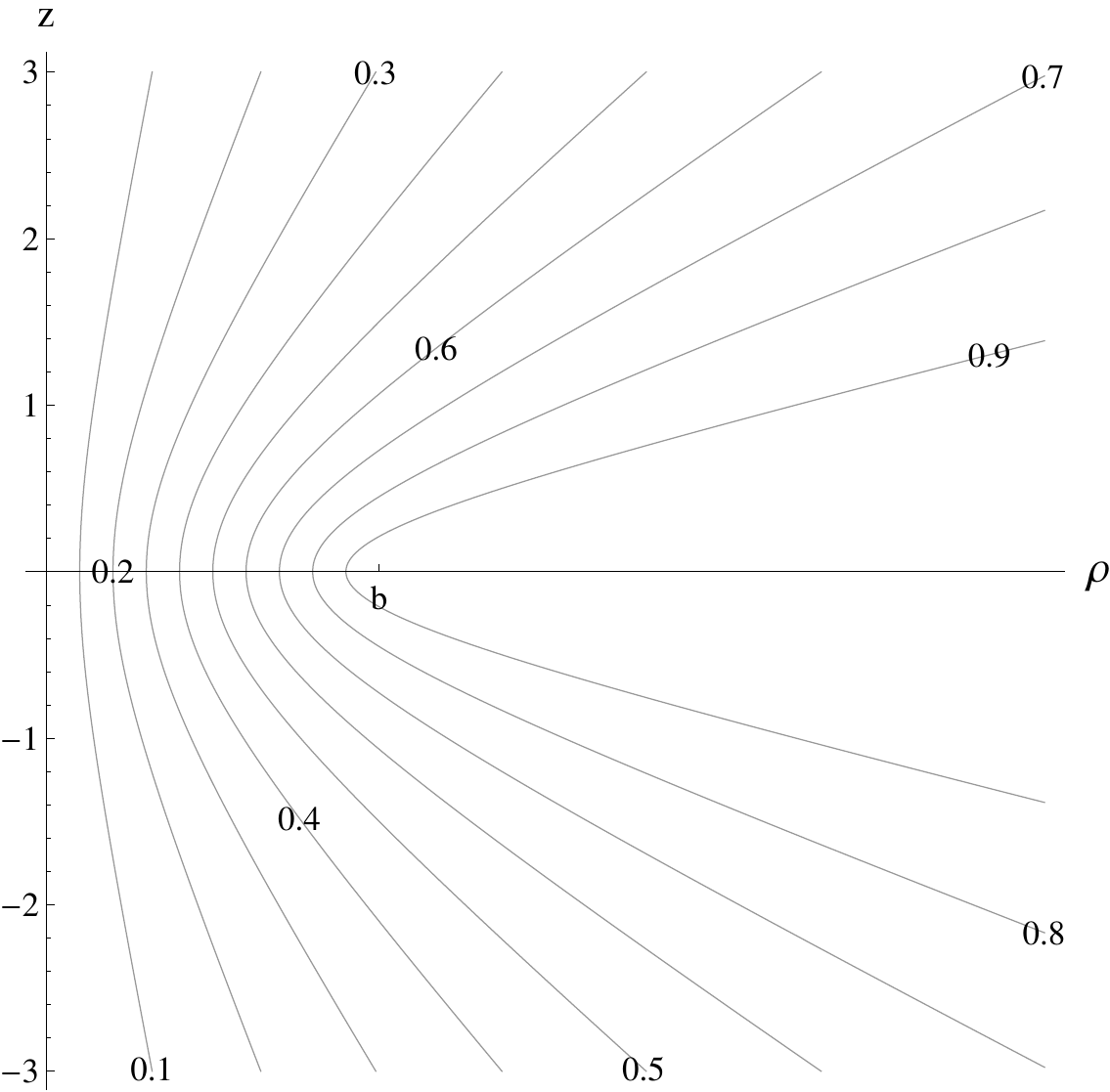}
	\caption{Level sets of $u(\rho,z)$. The coordinate $u$ ranges from $u=0$ on the axis $\theta=0$ to $u=1$ on the disk $\theta=\pi/2, \ \rho>b$. Each line $u\neq1$ meets the equator $z=0$ at $\rho=bu$. }
	\label{fig:uLevelSets}
\end{figure}

Our choice of $u(\rho,z)$ is derived from the confocal ellipsoidal coordinate system for flat space. The coordinate lines are ellipses and hyperbolae with foci at a distance $b$ from the origin. The curves of interest to us are the hyperbolae that arise as level sets of
\begin{align}
	\label{eq:EllipsoidalCoordinate}
	u=\frac{\sqrt{\pa{\rho+b}^2+z^2}-\sqrt{\pa{\rho-b}^2+z^2}}{2b},
\end{align}
which have a shape that could plausibly resemble the field lines in an AGN. Figure~\ref{fig:uLevelSets} plots these hyperbolae along with the corresponding value of $u\in[0,1]$. Plugging Eqs.~\eqref{eq:Unicorn} and \eqref{eq:EllipsoidalCoordinate} into Eq.~\eqref{eq:StreamEquation}, we are able to satisfy the equation provided that
\begin{align}
	\psi'(u)=\psi_0\frac{u}{\sqrt{\pa{1-u^2}\br{1-b^2\Omega(u)^2u^4}}},\qquad
	I(u)=\pm\frac{\psi_0}{b}\sqrt{\frac{1+b^2c\br{1-b^2\Omega(u)^2u^4}}{1-b^2\Omega(u)^2u^4}},
\end{align}
where $\psi_0$ and $c$ are constants and $\Omega(u)$ is an arbitrary function. The choice of sign corresponds to the direction of current and energy flow. In order to avoid a line current on the symmetry axis, $I(u)$ must vanish there, $I(0)=0$. This requires $c=-b^{-2}$, leading to 
\begin{align}
	\label{eq:Solution}
	\psi=\psi_0\int\frac{u\ed u}{\sqrt{\pa{1-u^2}\br{1-b^2\Omega(u)^2u^4}}},\qquad
	I=\pm\psi_0\frac{\Omega(u)u^2}{\sqrt{1-b^2\Omega(u)^2u^4}},
\end{align}
where the constant in the integral is to be fixed by requiring that $\psi(0)=0$ to make $\psi$ the flux function. For $\Omega=0$, we can perform the integral explicitly, resulting in
\begin{align}
	\label{eq:SolutionOmega0}
	\psi\big|_{\Omega=0}=\psi_0\pa{1-\sqrt{1-u^2}}.
\end{align}
This vacuum solution of Maxwell's equations was found previously in Ref.~\cite{Beskin:1992}. For $\Omega$ a nonzero constant we can represent the integral with elliptic functions (see Appendix~\ref{sec:EllipticFunctions}), while for arbitrary $\Omega(u)$ we can proceed no further. However, the electric and magnetic fields are directly expressible in terms of $\psi'(u)$, and therefore we have an explicit closed-form solution,
\begin{align}
	\label{eq:F}
	F=\frac{(\psi_0/2\pi)}{\sqrt{1-b^2\Omega(u)^2u^4}}
	\pa{\frac{u\ed u}{\sqrt{1-u^2}}\wedge\br{\ed\phi-\Omega(u)\ed t}
		\pm\Omega(u)\frac{u^2}{\rho}\ed z\wedge\ed\rho}.
\end{align}
Note that $u(\rho,z)$ depends on $b$ via \eqref{eq:EllipsoidalCoordinate}. Equation~\eqref{eq:F}, or equivalently Eq.~\eqref{eq:Solution}, provides a new exact family of force-free magnetospheres, parametrized by disk termination radius $b$, field line angular velocity $\Omega(u)$, and current/energy flow direction $\pm$.\footnote{One way to compute the electric and magnetic fields associated with the solution is to begin with \eqref{eq:Solution} and use the formulas given in footnote \ref{Footnote}.}

\subsection{Properties}

In order for the prefactor $\br{1-b^2\Omega(u)^2u^4}^{-1/2}$ in the solution to be finite and real, each field line $u$ must satisfy $bu^2<R_L(u)$, where $R_L(u)\equiv\Omega(u)^{-1}$ is the cylindrical radius of the light surface. Since $\rho=bu$ is the place where the field intersects the equator and $u\leq1$, the condition is satisfied if all field lines rotate subluminally on the equator, and in particular if they are anchored there to a rotating conductor (see the discussion in Sec.~\ref{sec:Star}).

The apparent singularity at the disk $u=1$ is just a discontinuity in physical coordinates (e.g., cylindrical), and corresponds to the presence of surface charge and current. In particular, the pullback of $F$ to the disk is continuous, indicating the absence of magnetic monopoles, but the pullback of the dual $\star F$ is discontinuous, revealing the presence of surface charge and current. Using the formulas (A23) and (A24) from Ref.~\cite{Gralla:2014yja}, we compute the surface charge and current to be
\begin{align}
	J_{\rm surf}=\mathcal{J}\delta(z)H(\rho-b)\ed z\wedge\ed\rho\wedge\pa{\ed t-\rho^2\,\Omega_1\ed\phi},\qquad
	\mathcal{J}=\frac{(\psi_0/\pi)}{\rho\sqrt{\pa{\rho^2-b ^2}\pa{1-b^2\Omega_1^2}}},
\end{align}
where $\delta(x)$ is the Dirac delta and $H(x)$ denotes the Heaviside function (vanishing for $x<0$ and unity for $x>0$). This corresponds to a surface current density $\vec{J}_{\rm surf}$ and charge density $\rho_{\rm surf}$ of
\begin{align}
	\vec{J}_{\rm surf}=\mathcal{J}\delta(z)H(\rho-b)\,\hat{\phi},\qquad
	\rho_{\rm surf}=\rho\,\Omega_1\ab{\vec{J}_{\rm surf}}.
\end{align}
Here, $\Omega_1\equiv\Omega\big|_{u=1}$ is the value of $\Omega$ as the field lines collapse the disk, which would normally be zero in physical models (e.g., those of Secs.~\ref{sec:Star} and \ref{sec:BH}). In this case, there is no surface charge, and the surface current is given by
\begin{align}
	\vec{J}_{\rm surf}=\frac{(\psi_0/\pi)}{\rho\sqrt{\rho^2-b^2}}\delta(z)H(\rho-b)\,\hat{\phi}.
\end{align}
The current density blows up at the disk edge $\rho=b$ (from the right), but the blowup grows only like the square root of the distance, so the total current is finite. Orthonormal frame components of the magnetic field also blow up like the square root of distance. This divergence is milder than the inverse-distance singularity that would be produced by a line current. Thus, our model contains a concentration of current near the edge of the disk, but not as much concentration as a line current. Similar comments apply to the charge and current density when $\Omega_1\neq0$.

Using Eq.~\eqref{eq:Flux}, the power (energy flux) flowing upward between two field lines $u_1<u_2$ is given by
\begin{align}
	\label{eq:Power}
	P_{\rm up}=\mp\frac{\psi_0^2}{2\pi}\int_{u_1}^{u_2}
		\frac{\Omega(u)^2u^3\ed u}{\sqrt{1-u^2}\br{1-b^2\Omega(u)^2u^4}}. 
\end{align}
We see that if the lower sign is chosen, the solution represents energy flowing in from infinity in the southern hemisphere and out to infinity in the northern, while the upper entails the opposite. For physical models of unipolar induction (Secs.~\ref{sec:Star} and \ref{sec:BH}), we have the freedom to choose the sign separately in each hemisphere for field lines that intersect the object and will do so in order to ensure outgoing flux everywhere.

In spherical coordinates,\footnote{The transformation to spherical coordinates is defined by $\rho=r\sin\theta$ and $z=r\cos\theta$.} the asymptotic $r\rightarrow\infty$ behavior of the function $u$ defined in Eq.~\eqref{eq:EllipsoidalCoordinate} is
\begin{align}
	\label{eq:uAsymptotic}
	u=\sin{\theta}+\mathcal{O}\!\pa{\frac{b^2}{r^2}}.
\end{align}
Thus, from Eq.~\eqref{eq:Solution}, the large-$r$ behavior of the solution is
\begin{align}
	\label{eq:AsymptoticCurrent}
	I&=\pm\psi_0\frac{\Omega\sin^2{\theta}}{\sqrt{1-b^2\Omega^2\sin^4{\theta}}}
		+\mathcal{O}\!\pa{\frac{1}{r^2}},\\
	\label{eq:AsymptoticFluxDerivative}
	\pd_\theta\psi&=\psi_0\frac{\sin{\theta}}{\sqrt{1-b^2\Omega^2\sin^4{\theta}}}
		+\mathcal{O}\!\pa{\frac{1}{r^2}}.
\end{align}
Since $\psi$ is asymptotically $r$-independent, the field lines are asymptotically radial. They are uniformly distributed (monopolar) when $b\Omega=0$ and become more bunched near the equator as this quantity increases. From Eqs.~\eqref{eq:AsymptoticCurrent} and \eqref{eq:AsymptoticFluxDerivative}, we see that\begin{align}
	\label{eq:AsymptoticRelation}
	I=\pm\Omega\sin{\theta}\pd_\theta\psi,\qquad r\rightarrow\infty.
\end{align}
This indicates that the electric and magnetic fields are asymptotically equal in magnitude.

The solution includes the radial and vertical solutions as special cases. The radial solution emerges as $b\rightarrow0$ at fixed spherical coordinates $r$ and $\theta$. From Eq.~\eqref{eq:uAsymptotic}, we have $u=\sin{\theta}+\mathcal{O}\!\pa{b^2}$, and hence from Eq.~\eqref{eq:F}, we find
\begin{align}
	\lim_{b\rightarrow0}F=\frac{\psi_0}{2\pi}\sin{\theta}\pa{\frac{\cos{\theta}}{\ab{\cos{\theta}}}\ed\theta\wedge
		\br{\ed\phi-\Omega(\sin{\theta})\ed t}\pm\Omega(\sin{\theta})\ed r\wedge\ed\theta}.
\end{align}
This is Michel's radial solution \cite{Michel:1973}, already in ``split'' form on account of the factor of $\cos{\theta}/\ab{\cos{\theta}}$. Normally, one derives the true monopole solution and then adds such a factor by hand. Here, we see the factor naturally emerge when the solution is regarded as a limiting case of the hyperbolic solution. Note that if one chooses $\pm$ oppositely in each hemisphere so as to have pure outgoing (or pure ingoing) flux, then the solution becomes proportional to $\sin{\theta}\ed\theta\wedge\br{\ed\phi-\Omega(\ed t\pm\ed r)}$ with opposite signs in opposite hemispheres, the simple form discussed in Ref.~\cite{Gralla:2014yja}.

The vertical solution emerges as $b\rightarrow\infty$ at fixed cylindrical coordinate. From Eq.~\eqref{eq:EllipsoidalCoordinate}, we have $u=\rho/b+\mathcal{O}\!\pa{b^{-3}}$, and hence from Eq.~\eqref{eq:F},
\begin{align}
	\label{eq:VerticalSolution}
	\lim_{b\rightarrow\infty}\pa{b^2F}=\frac{\psi_0}{2\pi}\rho\ed\rho\wedge\br{\ed\phi-\Omega(\ed t\pm\ed z)}.
\end{align}
In this particular limit, $\Omega$ is a constant because $\Omega(u\approx\rho/b)\rightarrow\Omega(0)$, but one may choose $\Omega$ to scale with $b$ so that the more general solution, Eq.~\eqref{eq:VerticalSolution} with $\Omega=\Omega(\rho)$, is obtained. Time-dependent, nonaxisymmeric generalizations of vertical solutions are given in Ref.~\cite{Gralla:2015wva}.

\section{Star with a Disk}
\label{sec:Star}

Unipolar induction (also homopolar induction) is the generation of an electric field by a conductor moving in a magnetic field. The effect may be understood directly from the perfect conductor assumption that the electric field must vanish in the rest frame of the conductor. If the conductor moves with velocity $\vec{v}$ (and boost factor $\gamma$) in a magnetic field $\vec{B}$, the rest-frame electric field is given by $\gamma(\vec{E}+\vec{v}\times\vec{B})$, so that an electric field of $-\vec{v}\times\vec{B}$ must be generated. If wires (or plasma) are connected to the conductor, then the resulting voltage will drive a current, and energy will be carried away. In the case of an ordinary circuit, the amount of current and power is fixed by the resistance of the load. In the case of a conductor in plasma, one must self-consistently solve the equations, from which an effective resistance can be determined.

In the framework of stationary, axisymmetric force-free magnetospheres, the boundary condition of a rotating conductor is simple and intuitive: field lines meeting the conductor must rotate with it. This may be seen by recalling that the electric field in a frame with four-velocity $u^\alpha$ is given by $F_{\alpha\beta}u^\beta$. For a conductor rotating with angular velocity $\omega$, the four-velocity (field) is $u^\alpha=\gamma(1,0,0,\omega)$ in cylindrical coordinates. Plugging into Eq.~\eqref{eq:CylindricalAnsatz}, we see that $F_{ab}u^b=0$ requires that $\Omega(\psi)=\omega$ for any field line $\psi$ that intersects the conducting surface.\footnote{In general, only the tangential components of the electric field must be continuous across a surface layer, whereas here we have demanded that all components be continuous (and hence vanish). If one considers only the tangential components, then $\Omega=\omega$ is still required provided that the field line intersects the conductor nontangentially. Furthermore, this makes all components vanish, so that there is no induced surface charge in the conductor frame. For further discussion, see Sec.~8.1 of Ref.~\cite{Gralla:2014yja}.} This leads to the intuitive picture of field lines being anchored to the conductor.

In light of the preceding discussion, it is straightforward to use our solution to model unipolar induction by a star of radius $R<b$ rotating at constant angular velocity $\Omega_\star$ in the field of a magnetized disk. We use the force-free solution outside the star and demand that field lines meeting the star corotate with it,
\begin{align}
	\Omega(u)=\Omega_\star H(R-bu),
\end{align}
where $H(x)$ again denotes the Heaviside function. Second, we must impose the boundary condition that no energy comes in from infinity, by selecting the lower sign in the northern hemisphere and the upper sign in the southern hemisphere, $\pm\rightarrow-{\rm sign}(\theta-\pi/2)$ in Eq.~\eqref{eq:F} [and Eq.~\eqref{eq:Solution}]. Thus, the self-consistent model of unipolar induction by a rotating star in plasma in a magnetic field provided by a thin disk is 
\begin{align}
	F=\frac{(\psi_0/2\pi)}{\sqrt{1-b^2\Omega(u)^2u^4}}\pa{\frac{u\ed u}{\sqrt{1-u^2}}\wedge
		\br{\ed\phi-\Omega_\star H(R-bu)\ed t}-s\,\Omega_\star H(R-bu)\frac{u^2}{\rho}\ed\rho\wedge\ed z},
\end{align}
where
\begin{align}
	s\equiv{\rm sign}\!\pa{\theta-\frac{\pi}{2}}.
\end{align}
The model is symmetric under reflection about the equatorial plane, and for simplicity, we shall henceforth consider solely the northern hemisphere, $0\leq\theta\leq\pi/2$. The flux function is given by\footnote{Since the star cannot rotate faster than the speed of light, we have $R\Omega_\star<1$ and hence $b\Omega u<1$, making $b^2\Omega^2u^4\ll1$ when $R\ll b$.}
\begin{align}
	\psi=\psi_0\int\frac{u\ed u}{\sqrt{\pa{1-u^2}\br{1-b^2\Omega_\star^2u^4H(R-bu)}}}
	\approx\psi_0(1-\sqrt{1-u^2}),\qquad R\ll b.
\end{align}
Thus the total magnetic flux on the star is
\begin{align}
	\label{eq:TotalStarFlux}
	\Psi_\star=\psi_0 \int_{0}^{R/b}\frac{u\ed u}{\sqrt{\pa{1-u^2}\br{1-b^2\Omega_\star^2u^4H(R-bu)}}}
	\approx\frac{\psi_0}{2}\pa{\frac{R}{b}}^2,\qquad R\ll b.
\end{align}
If $R\not\ll b$, these integrals are given by elliptic functions (Appendix~\ref{sec:EllipticFunctions}).

\subsection{Star Jet Properties} 

The model contains current and energy flux only within the last field line touching the star, $u\leq R/b$. Since the field lines are asymptotically radial, this may be called a conical jet. In spherical coordinates, we have $u\approx\sin{\theta}$ for large $r$ from Eq.~\eqref{eq:uAsymptotic}, so the (half-)opening angle is just 
\begin{align}
	\label{eq:StarJetAngle}
	\theta^{\rm \star}_{\rm jet}=\arcsin{\frac{R}{b}}
	\approx\frac{R}{b},\qquad R\ll b .
\end{align}
From Eqs.~\eqref{eq:Power} and \eqref{eq:uAsymptotic}, the outgoing power per unit angle $\theta$ in the asymptotic jet is
\begin{align}
	\label{eq:StarJetPower}
	\frac{dP}{d\theta}=\frac{\psi_0^2}{2\pi}
		\frac{\Omega_\star^2\sin^3{\theta}}{1-b^2\Omega_\star^2\sin^4{\theta}},\qquad r\to\infty,
\end{align}
which is concentrated toward the edge of the cone. (This formula includes only the contribution from one hemisphere; the total power is a factor of 2 larger.) Recall that $R_L(u)\equiv\Omega(u)^{-1}$ and hence $b\Omega_\star=b/R_L$ is the ratio of the disk radius to the light cylinder radius.

Using Eq.~\eqref{eq:StarJetPower}, we can compute the total power from a slowly rotating star ($b\Omega_{\star}\ll1$),
\begin{align}
	P_{\rm jet}^\star=\frac{\psi_0^2\,\Omega_\star^2}{2\pi}\int_0^{\theta^{\rm \star}_{\rm jet}}\sin^3{\theta}
	=\frac{\psi_0^2\,\Omega_\star^2}{3\pi}\br{1-\sqrt{1-\frac{R^2}{b^2}}\pa{1+\frac{R^2}{2b^2}}}.
\end{align}
If in addition $R\ll b$, then
\begin{align}
	 P_{\rm jet}^\star\approx\frac{\psi_0^2\,\Omega_\star^2}{8\pi}\pa{\frac{R}{b}}^4
	 =\frac{\pa{\Psi_\star\Omega_\star}^2}{2\pi}.
\end{align}

\subsection{Circuit Model}

The net current flowing in the bulk of the jet is given by $I(u=R/b)$,
\begin{align}
	I_{\rm bulk}=-\psi_0\frac{\Omega_\star(R/b)^2}{\sqrt{1-(R\Omega_\star)^2(R/b)^2}},
\end{align}
while an equal and opposite remainder flows along the current sheet at $u=R/b$. [This ``return flow'' is required by current conservation and may also be computed directly from the jump in $\star F$ caused by the discontinuous choice of $\Omega(u)$.] The voltage drop from the pole to the equator (or equivalently, to the end of the jet) is just $\Omega_\star\psi\big|_{u=R/b}$ and hence can be expressed in terms of elliptic functions (see Appendix~\ref{sec:EllipticFunctions}). To leading order in $b\Omega_\star$, it is simply
\begin{align}
	V=\Omega_\star\psi\big|_{u=R/b}
	=\frac{\psi_0\,\Omega_\star}{2\pi}\br{1-\sqrt{1-(R/b)^2}}+\mathcal{O}\!\pa{b^2\Omega_\star^2}.
\end{align}
This is the approximation that the light cylinder radius $1/\Omega_\star$ is well outside the disk edge $b$. Since $R<b$, this approximation also entails $R\Omega_\star\ll1$ small as well, i.e., that the surface velocity of the star is much less than the speed of light. Taking this into account, we have for the effective resistance $R_{\rm eff}$ of the plasma in the circuit
\begin{align}
	\label{eq:StarResistance}
	R_{\rm eff}=\frac{V}{\ab{I_{\rm bulk}}}
	\approx\frac{1-\sqrt{1-(R/b)^2}}{(R/b)^2},\qquad b\Omega_\star\ll1.
\end{align}
Here, $R_{\rm eff}$ varies from $1/2$ to $1$ as $R/b$ varies from $0$ to $1$. The value $R_{\rm eff}=1$ is the ``impedance of free space'' ($4\pi$ in cgs-Gaussian units), often used as the effective resistance of the plasma for rough estimates (e.g., Ref.~\cite{Lai:2012qe}). Thus, for our setup in the slow rotation approximation, the true effective resistivity of the plasma differs from that of ``free space'' by at most a factor of 2. For rapidly rotating stars, one may use the expressions in Appendix~\ref{sec:EllipticFunctions}. The effective resistance never exceeds $1$ and changes order of magnitude only when both $\Omega_\star R$ and $R/b$ are near 1, where it drops to zero. This is the limit where the star nearly touches the disk and has surface velocity nearly equal to the speed of light.

If we instead assume $R/b\ll1$, then from Eq.~\eqref{eq:TotalStarFlux}, we have $V=\Omega_\star\Psi_\star=\tfrac{1}{2}\psi_0\,\Omega_\star(R/b)^2$. The effective resistance is thus
\begin{align}
	R_{\rm eff}=\frac{1}{2},\qquad R\ll b.
\end{align}
This equation does {\it not} assume $b\Omega_\star\ll1$ nor $R\Omega_\star\ll1$. 

\section{Black Hole with a Disk}
\label{sec:BH}

In their original paper \cite{Blandford:1977ds}, Blandford and Znajek derived the stream equation in the Kerr spacetime [the analog of Eq.~\eqref{eq:StreamEquation}] and laid out a perturbative method for solving it. One begins by expanding the flux function, current, and angular velocity\footnote{In Kerr, or more generally any stationary axisymmetric (circular) spacetime, these quantities are invariantly defined in the same way as in flat spacetime \cite{Gralla:2014yja}.} in the spin of the black hole as follows,\footnote{The leading scalings follow from the fact that no energy can be extracted from a nonrotating black hole. [Note that Eq.~\eqref{eq:Flux} holds in Kerr.] The scaling of the error terms may be seen from the Znajek condition (e.g., Eq.~(104) of Ref.~\cite{Gralla:2014yja}).}
\begin{align}
	\label{eq:BlackHoleFlux}
	\psi&=X(r,\theta)+\mathcal{O}\!\pa{\frac{a^2}{M^2}},\\
	\label{eq:BlackHoleCurrent}
	I&=\frac{a}{M^2}\,Y(X)+\mathcal{O}\!\pa{\frac{a^3}{M^3}},\\
	\label{eq:BlackHoleOmega}
	\Omega&=\frac{a}{M^2}\,W(X)+\mathcal{O}\!\pa{\frac{a^3}{M^3}},
\end{align}
where $X$, $Y$, and $W$ are independent of the spin $a$.\footnote{Our notation differs slightly from BZ who use $A_{\phi}\equiv\psi-1=X+\mathcal{O}(a/M)^2$.} [We work in Boyer-Lindquist coordinates $(t,r,\theta,\phi)$.] Since the current and angular velocity vanish at $a=0$, the leading-order flux function $X$ must be a vacuum Maxwell solution in the Schwarzschild spacetime. At next order $\mathcal{O}(a)$, the equations are automatically satisfied by the form of Eqs.~\eqref{eq:BlackHoleFlux}--\eqref{eq:BlackHoleOmega}---the entire content of the force-free equations is the statement that $Y$ and $W$ are functions of $X$ alone. The Kerr force-free stream equation appears first at $\mathcal{O}\!\pa{a^2}$, which we do not consider in this paper.

To find force-free solutions to $\mathcal{O}(a)$, then, one needs a vacuum Schwarzschild solution $X$ as well as two further equations to fix $Y$ and $W$. The first comes from a universal relationship that always holds on the black hole horizon, the so-called Znajek condition \cite{Znajek:1977},
\begin{align}
	\label{eq:ZnajekHorizon}
	Y=\pa{W-\frac{1}{4}}\sin{\theta}\pd_\theta X\qquad\text{at }r=r_H,
\end{align}
which we present here to the relevant order in the spin. This condition guarantees that the fields are regular at the horizon and also follows from the stream equation (e.g., Ref.~\cite{Gralla:2014yja}). Note that $\tfrac{1}{4}$ is the value of $W$ corresponding to the horizon angular frequency. The second equation, according to a key idea of the BZ approach, is to be determined from an {\it analogous force-free solution in flat spacetime}. If the goal is simply to satisfy the force-free equations to $\mathcal{O}(a)$, then any postulated second relationship between $X$, $Y$, and $W$ will do. But this is an extremely large freedom and may correspond to spurious linearized solutions that arise from {\it no} family of exact solutions. In order to eliminate this freedom, BZ postulated that a second relationship should be determined from an exact flat spacetime solution that agrees with the linearized solution at large $r$, where the Kerr spacetime becomes approximately flat. In our case, this second relationship will be Eq.~\eqref{eq:AsymptoticRelation} with the lower sign appropriate to outgoing flux in the northern hemisphere,
\begin{align}
	\label{eq:ZnajekInfinity}
	Y=-W\sin{\theta}\pd_\theta X\qquad\text{as }r\rightarrow\infty.
\end{align}
Here and below, we work in the northern hemisphere $0\leq\theta\leq\pi/2$, with the southern fields determined by reflection.

\subsection{Flux Function $X$}

The first step of the BZ method is to find a vacuum ($I=\Omega=0$) flux function $X$ in Schwarzschild that corresponds to the ``same'' physical configuration as the flat spacetime hyperbolic force-free solution $\psi$ evaluated at $I=\Omega=0$. In the cases treated by BZ, there was an obvious ``same'' solution, but here we must exercise more care. In general, there is no canonical mapping between field configurations on different spacetimes, but in this case, we have just enough structure to make one. Using the facts that (i) the flux functions must satisfy the vacuum Maxwell equation in their respective spacetimes (flat and Schwarzschild) and (ii) these spacetimes both exhibit spherical symmetry, we can make a unique identification mode by mode in an expansion in angular harmonics. In Appendix~\ref{sec:MainAppendix}, we follow this general approach and derive a family of solutions parametrized by the Schwarzschild coordinate radius $b_\circ$ of the disk, expressed in series expansions valid for $r\leq b_\circ$ and $r\geq b_\circ$. The flux function for this family is given by Eqs.~\eqref{eq:SchwarzschildSolution}, \eqref{eq:b}, and \eqref{eq:C}.

For the discussion here, we need only the following properties of the solution:
\begin{align}
	\label{eq:HorizonX}
	X(r=r_H,\theta)&\approx X_H\sin^2{\theta},\\
	\label{eq:InfinityX}
	X(r\to\infty,\theta)&=X_0(1-\cos{\theta}).
\end{align}
These indicate that the field is approximately vertical on the black hole and approximately radial (monopolar) at infinity. We have multiplied the solution \eqref{eq:SchwarzschildSolution} by an overall constant $X_0$ representing the magnetic flux in one hemisphere. Equation~\eqref{eq:InfinityX} follows from \eqref{eq:InfinityFlux} as $r\rightarrow\infty$, noting that $R^>_\ell(r)\sim r^{-\ell}$. Equation~\eqref{eq:HorizonX} follows from \eqref{eq:HorizonFlux}, keeping only the leading $k=1$ term, noting that $R^<_1(r)=r^2$. Dropping the higher-order terms is an excellent approximation provided that the disk does not extend too close to the black hole. For example, for $b_\circ\geq6M$, the subleading term is at least a factor of $10^{-3}$ smaller. The constant $X_H$ is the total flux on one hemisphere of the black hole horizon and is given by
\begin{align}
	X_H=CX_0\frac{8M^2}{\br{b_\circ-M+\sqrt{b_\circ(b_\circ-2M)}}^2},
\end{align}
where $C(M,b_\circ)$ is a constant of order unity given in Eq.~\eqref{eq:C} [see also Eqs.~\eqref{eq:KummerExact}--\eqref{eq:KummerFirstTerm}]. For $b_\circ=6M$ (the ISCO radius), $C\approx.735$, while for $b_\circ\gg M$ (disk very far), $C=1$. Thus,
\begin{align}
	\label{eq:MatchingCoefficients}
	\left.\frac{X_H}{X_0}\right|_{b_\circ=6M}\approx.060,\qquad
	\left.\frac{X_H}{X_0}\right|_{b_\circ\rightarrow\infty}\approx\frac{2 M^2}{b_\circ^2}.
\end{align}
At $b_\circ=6M$, the $b_\circ\rightarrow\infty$ expression gives $X_H/X_0\approx0.056$, quite close to the true value of $0.060$.

\subsection{Current $Y$ and Angular Velocity $W$}

Plugging Eq.~\eqref{eq:HorizonX} into \eqref{eq:ZnajekHorizon} gives a condition at the horizon, while plugging Eq.~\eqref{eq:InfinityX} into \eqref{eq:ZnajekInfinity} gives a condition at infinity. These, respectively, are
\begin{align}
	Y&=2X_H\pa{W-\tfrac{1}{4}}\sin^2{\theta}\cos{\theta}&\text{at }r=r_H.\\
	Y&=-X_0W\sin^2{\theta}&\text{as }r\rightarrow\infty.
\end{align}
These equations hold at different radial coordinates and cannot yet be compared. However, we may use Eqs.~\eqref{eq:HorizonX} and \eqref{eq:InfinityX} to eliminate $\theta$ in favor of $X$,
\begin{align}
	Y&=2\pa{W-\tfrac{1}{4}}X\sqrt{1-X/X_H}&(\text{from the horizon; }X<X_H),\\
	\label{eq:MonopolarY}
	Y&=-WX(2-X/X_0)&(\text{from infinity; all } X).
\end{align}
Since $Y$ and $W$ are functions only of $X$, the equations may now be combined, giving
\begin{align}
	\label{eq:Y}
	Y(X)&=-W X(2-X/X_0),\\
	\label{eq:W}
	W(X)&=\frac{1}{4} \frac{\sqrt{1-X/X_H}}{1-X/(2X_0)+\sqrt{1-X/X_H}},
\end{align}
which hold on the overlapping domain of validity, $X<X_H$. For $X>X_H$, we take $Y=W=0$, as required by reflection symmetry and smoothness. Note that this does not entail any current sheet since $Y=W=0$ at $X=X_H$ from Eqs.~\eqref{eq:Y} and \eqref{eq:W}. Recall that $1/4$ is the value of $W$ corresponding to the horizon angular frequency. 

A similar procedure was performed previously by Ref.~\cite{Beskin:1992} (see also Ref.~\cite{Beskin:2010iba}), who arrived at Eqs.~\eqref{eq:Y} and \eqref{eq:W} in the distant-disk limit $b_\circ\gg M$ (equivalently $X_H\ll X_0$). In this limitm=, the terms with $X/X_0$ disappear, since $X<X_H$ in those expressions. They postulated Eq.~\eqref{eq:MonopolarY}, which we derived here from the corresponding flat force-free solution \eqref{eq:Solution} according to the BZ method. They found the vacuum flat solution \eqref{eq:FlatSpaceVacuumSolution} in generality and used it to find the near-horizon behavior of the Schwarzschild vacuum solution when $b_\circ\gg M$. (In this limit, $C=1$ and $b=b_\circ$, and the many subtleties discussed in Appendix~\ref{sec:MainAppendix} do not arise.) We have found the vacuum solution everywhere outside the horizon and for arbitrary $b_\circ$, an effort which reveals that the $b_\circ\rightarrow\infty$ expression in Eq.~\eqref{eq:MatchingCoefficients} in fact works well even at the ISCO radius $b_\circ=6M$. Equations~\eqref{eq:Y} and \eqref{eq:W} without the $X/X_0$ terms were also obtained in Ref.~\cite{Pan:2014bja} by considering strictly vertical field lines and imposing an integrability condition for the second-order perturbation.

\subsection{Black Hole Jet Properties}

As in the exact flat solution, the approximate Kerr solution produces a conical jet. The boundary of the jet is the last field line touching the black hole, labeled by $X=X_H$. From Eq.~\eqref{eq:InfinityX}, this has an asymptotic $\theta$-angle of
\begin{align}
	\label{eq:BlackHoleJetAngle}
	\theta^\bullet_{\rm jet}=\arccos\pa{1-\frac{X_H}{X_0}}
	\approx
	\begin{cases}
		20^\circ & \qquad b_\circ=6M,\\
 		\frac{2M}{b_\circ} & \qquad b_\circ\gg M.
	\end{cases}
\end{align}
At large $r$, the power per unit $\theta$ in the jet is
\begin{align}
	\frac{dP}{d\theta}=\frac{X_0^2\,\Omega_H^2}{2\pi}w(\theta)^2\sin^3{\theta},
\end{align}
where
\begin{align}
	w(\theta)=\frac{\sqrt{\cos{\theta}-\cos{\theta^\bullet_{\rm jet}}}}
		{\sqrt{1-\cos{\theta^\bullet_{\rm jet}}}\cos^2{\tfrac{\theta}{2}}
		+\sqrt{\cos{\theta}-\cos{\theta^\bullet_{\rm jet}}}},
\end{align}
and $\Omega_H=a/(4M^2)$ at this order of approximation.
 
The total power may also be computed from Eq.~\eqref{eq:Flux}, which holds in any stationary, axisymmetric, circular spacetime \cite{Gralla:2014yja}. We can compute numerically for all finite values of $b_\circ$ or get an analytic expression when $b_\circ\gg M$ (equivalently $X_H\ll X_0$),
\begin{align}
	P^{\bullet}_{\rm jet}\approx
	\begin{cases}
		7\times10^{-5}\,X_0^2\,\Omega_H^2\approx.02\,X_H^2\,\Omega_H^2 & \qquad b_\circ=6M,\\
		\displaystyle\frac{\Omega_H^2}{\pi}\int_0^{X_H}X\pa{\tfrac{\sqrt{1-X/X_H}}{1+\sqrt{1-X/X_H}}}^2\ed X
		=\frac{17-6\log{16}}{6\pi}X_H^2\,\Omega_H^2 & \qquad b_\circ\gg M,
	\end{cases}
\end{align}
where we have used Eq.~\eqref{eq:MatchingCoefficients}. Note that $(17-6\log 16)/(6\pi)\approx.02$, so that the large-$b_\circ$ expression in fact gives an excellent approximation for the $b_\circ=6M$ answer. These expressions for the power include only one jet; the total power is a factor of 2 larger.

\subsection{Circuit Model}

The voltage drop across the jet is the integral of $\Omega\ed\psi$. We can compute this analytically in the limit $b_\circ\gg M$ (equivalently $X_H\ll X_0$),
\begin{align}
	V=\int\Omega\ed\psi=\Omega_H\int_0^{X_H}\frac{\sqrt{1-X/X_H}}{1+\sqrt{1-X/X_H}}\ed X
	=(\log{4}-1)X_H\,\Omega_H,\qquad b_\circ\gg M .
\end{align}
In this limit, the net outward current $-I$ reaches a maximum value of $\tfrac{1}{2}X_H\,\Omega_H$ at $X=\tfrac{3}{4}X_H$. We regard $X<\tfrac{3}{4}X_H$ as the ``bulk'' flow and $X>\tfrac{3}{4}X_H$ as the return flow and write
\begin{align}
	I_{\rm bulk}=-\frac{X_H\,\Omega_H}{2}.
\end{align}
This is the current in the circuit, and hence the effective resistance is
\begin{align}\label{eq:BlackHoleResistance}
	R_{\rm eff}=\frac{V}{\ab{I_{\rm bulk}}}=2(\log{4}-1)\approx.77.
\end{align}
This analysis holds for $b_\circ\gg M$ but provides a good approximation even for $b_\circ=6M$.

\section{Stars vs. Black Holes}
\label{sec:StarBH}

\begin{figure*}
	\centering
	\subfigure[Angular Frequency]{\includegraphics[width=0.45\textwidth]{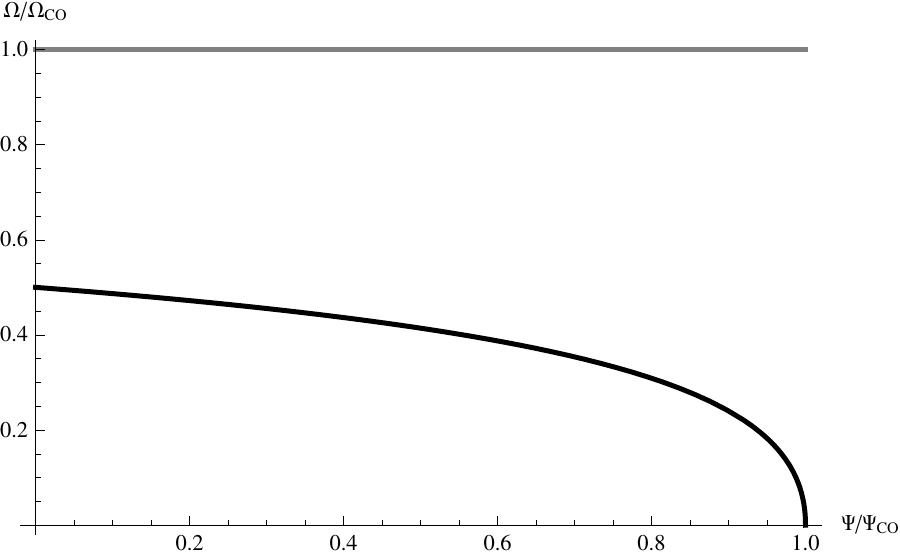}}
	\hspace{1cm}
	\subfigure[Current Enclosed]{\includegraphics[width=0.45\textwidth]{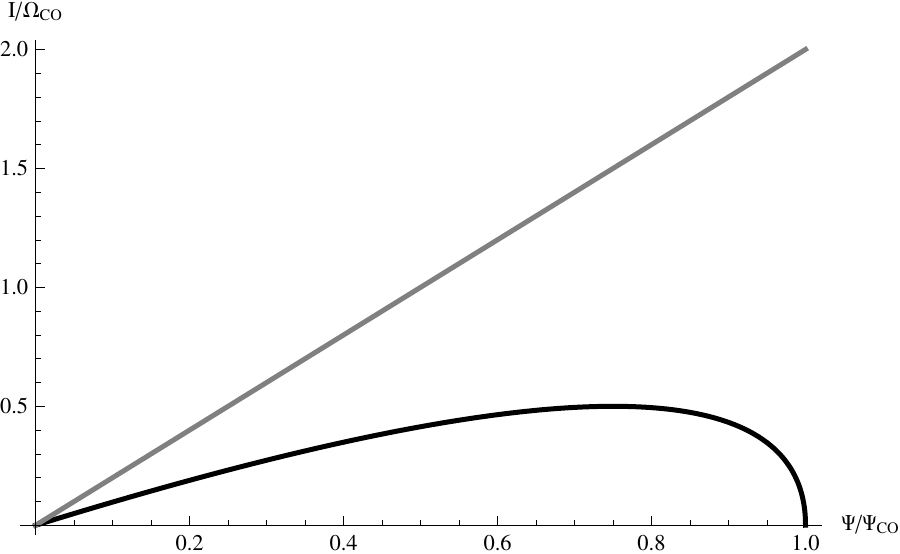}} 
	\caption{Comparison of jets from stars and black holes immersed in the magnetic field of a thin disk. On the left, field line angular velocity as a function of magnetic flux, normalized by the compact object velocity $\Omega_{\rm CO}$ and the magnetic flux on the compact object $\Psi_{\rm CO}$, respectively. In the star case, the field angular velocity equals that of the star, while in the black hole case, it is variable, peaking at half the black hole frequency on the pole and vanishing on the equator. The star current and field velocity do not vanish at the edge of the jet, indicating the presence of a current sheet there. To make a direct comparison, we have assumed that the compact objects are slowly spinning and that the the disk is far away. More generally, the current in the star's jet is a nonlinear function of magnetic flux, as indicated in Eq.~\eqref{eq:Solution}.}
	\label{fig:StarBH}
\end{figure*}

We conclude with a side-by-side comparison of the star (Sec.~\ref{sec:Star}) and black hole (Sec.~\ref{sec:BH}) solutions. The main qualitative difference is that the star's jet is surrounded by a current sheet, while the black hole's jet is not. To compare more quantitatively, we must specify how the parameters are to be chosen in each case, make the same approximations in both compact object (CO) solutions, and compare only invariantly defined quantities. On the one hand, the star solution is parametrized by radius $R$, disk edge $b$, and angular velocity $\Omega_\star$. On the other hand, the black hole solution is parametrized by black hole mass $M$ (also proportional to its size), Schwarzschild coordinate disk edge $b_\circ$, and angular velocity $\Omega_H$. We will take $\Omega_{\rm CO}\equiv\Omega_H=\Omega_\star$ and assume that the disk edge is far from the compact object, $b\gg R$ and $b_\circ\gg M$; then, we may take $b_{\rm CO}\equiv b=b_\circ$ without any coordinate ambiguity. There is no unambiguous way to compare the stellar radius $R$ with the black hole mass $M$. Instead, we will demand that the opening angles of the jets are the same. From Eqs.~\eqref{eq:StarJetAngle} and \eqref{eq:BlackHoleJetAngle}, we have $\theta^\star_{\rm jet}=R/b$ and $\theta^\bullet_{\rm jet}={2M}/{b_\circ}$. Hence, setting them equal gives
\begin{align}
	\theta^\bullet_{\rm jet}=\theta^\star_{\rm jet}\qquad\Longrightarrow\qquad R=2M,
\end{align}
which happens to be the Schwarzschild coordinate radius of the event horizon. This also makes the total magnetic flux $\Psi_{\rm CO}\equiv\Psi_\star=X_H=\frac{1}{2}\psi_0(R/b)^2$ on the objects equal---see Eqs.~\eqref{eq:TotalStarFlux} and \eqref{eq:MatchingCoefficients}---while identifying $X_0=\psi_0$. With these identifications, we have
\begin{align}
	P^\star_{\rm jet}=\frac{\pa{\Psi_{\rm CO}\Omega_{\rm CO}}^2}{2\pi},\qquad
	P^\bullet_{\rm jet}=\frac{17-6\log{16}}{6\pi}\pa{\Psi_{\rm CO}\,\Omega_{\rm CO}}^2
	\qquad\Longrightarrow\qquad
	P^\star_{\rm jet}\approx10\times P^\bullet_{\rm jet}.
\end{align}
The total power of the star is approximately ten times larger than the black hole's power. In this limit, we have the simple expressions \eqref{eq:StarResistance} and \eqref{eq:BlackHoleResistance} for the effective resistance of the circuit, 
\begin{align}
	R^\star_{\rm eff}=.5,\qquad R^\bullet_{\rm eff}=2(\log{4}-1)\approx.77. 
\end{align}
In analytical estimates, one sometimes assigns the impedance of free space $R_{\rm free}=1\,(\approx 377$ Ohms) to the ambient space and to the black hole. This would suggest $R^\star_{\rm eff}=1$ and $R^\bullet_{\rm eff}=2$, which is off by a factor of about 2. However, the basic intuition that the black hole provides extra resistance is supported by the exact solution.

Since the black hole solution is valid only for $M\Omega_H\ll1$, we should similarly assume $R\Omega_\star\ll1$ in the star case. With all these approximations, Fig.~\ref{fig:StarBH} plots the angular velocity and current functions of the star and the black hole. The physical setup is nearly identical, and the compact objects are effectively vanishingly small in the approximations taken. Nevertheless, the distant jet has dramatically different properties in each case. In this way, the force-free magnetosphere carries information about the nature of the central object to observers located at infinity.

\section*{Acknowledgements}

This work was supported in part by NSF grants No.~1205550 and No.~1506027 and the Fundamental Laws Initiative at Harvard. We thank Ramesh Narayan for helpful conversations.

\appendix

\section{Flux Function for Constant $\Omega$}
\label{sec:EllipticFunctions}
 
We now evaluate the integral in Eq.~\eqref{eq:Solution} for $\psi$ in the case that $\Omega(u)=\Omega$ is a nonzero constant. We must then require that $\ab{b\Omega}\in(0,1)$ and find
\begin{align}
	\psi(u)=\frac{\psi_0}{\sqrt{b\Omega\pa{b\Omega+1}}}
		\mathcal{E}\!\pa{\arcsin{\sqrt{\frac{b\Omega u^2+1}{2}}},\frac{2}{b\Omega+1}}+D,
\end{align}
where $\mathcal{E}(\phi,m)$ is the elliptic integral of the first kind. The integration constant $D$ must be fixed such that $\psi$ vanishes on the pole, $\psi(0)=0$. This gives
\begin{align}
	\psi(u)=\frac{\psi_0}{\sqrt{b\Omega\pa{b\Omega+1}}}
	\br{\mathcal{E}\!\pa{\arcsin{\sqrt{\frac{b\Omega u^2+1}{2}}},\frac{2}{b\Omega+1}}
		-\mathcal{E}\!\pa{\frac{\pi}{4},\frac{2}{b\Omega+1}}}.
\end{align}

\section{Vacuum Solution in Schwarzschild}
\label{sec:MainAppendix}

In this seven-part appendix we discuss general procedures for identifying vacuum (source-free) Maxwell solutions in flat spacetime with corresponding solutions in Schwarzschild spacetime and present an explicit construction for the hyperbolic case of relevance to this paper.

\subsection{Overview}
\label{sec:Overview}

In the vacuum case ($I=\Omega=0$), the stream equation \eqref{eq:StreamEquation} in flat spacetime reduces to
\begin{align}
	\label{eq:FlatSpaceVacuumEquation}
	r^2\pd_r^2\psi+\pd_\theta^2\psi-\cot{\theta}\pd_\theta\psi=0,
\end{align}
where now we work in spherical coordinates $(r,\theta)$. The corresponding equation in Schwarzschild spacetime (with Schwarzschild coordinates) is
\begin{align}
	\label{eq:SchwarzschildVacuumEquation}
	r(r-2M)\pd_r^2\psi+2M\pd_r\psi+\pd_\theta^2\psi-\cot{\theta}\pd_\theta\psi=0,
\end{align}
where $M$ is the black hole mass.\footnote{Equations~\eqref{eq:FlatSpaceVacuumEquation} and \eqref{eq:SchwarzschildVacuumEquation} are equivalent to the vacuum Maxwell equations, in their respective spacetimes, for an axisymmetric vector potential $A_\mu$ in a gauge where only $A_\phi=\psi$ is nonvanishing. They differ from the equations for the magnetostatic scalar potential used in Ref.~\cite{Anderson:1970} and many other references.\label{FootnoteBis}} These equations are separable, have the same angular part at all $r$, and agree entirely as $r\rightarrow\infty$. These facts allow one to make a canonical identification between solutions in the $M=0$ and $M\neq0$ cases, as we now explain.

A physical flux function $\psi$ must vanish at both poles. As discussed in Appendix~\ref{sec:Harmonics} below, a sufficiently regular solution of \eqref{eq:SchwarzschildVacuumEquation} vanishing at both poles can be expanded, at least locally in $r$, as
\begin{align}
	\label{eq:SeriesExpansion}
	 \psi(r,\theta)=\sum_{\ell=1}^\infty\br{C^<_\ell R^<_\ell(r)+C^>_\ell R^>_\ell(r)}\Theta_\ell(\theta),
\end{align}
where $C^<_\ell$ and $C^>_\ell$ are (real) constants. The angular harmonics $\Theta_\ell$ vanish at the poles, while $R^<_\ell$ is regular at the horizon (but not at infinity) and $R^>_\ell$ is regular at infinity (but not at the horizon). Solutions of the $M=0$ equation \eqref{eq:FlatSpaceVacuumEquation} admit an expansion analogous to that in Eq.~\eqref{eq:SeriesExpansion}, with identical angular harmonics $\Theta_{\ell}(\theta)$. The radial harmonics are different but may be identified with corresponding $M\neq0$ radial harmonics by demanding that they agree as $r\rightarrow\infty$, where the equations agree. Thus, one may associate solutions of Eqs.~\eqref{eq:FlatSpaceVacuumEquation} and \eqref{eq:SchwarzschildVacuumEquation} by expressing them in the form \eqref{eq:SeriesExpansion} and demanding that the coefficients $C^\gtrless_\ell$ be the same.\footnote{While this can always be implemented in a sufficiently small (but finite) domain, in typical problems, $C^<_\ell$ and $C^>_\ell$ will be chosen differently on different domains (e.g., matched at some radius). In this case, there remains the ambiguity of how to choose the domains. This ambiguity does not arise in the related method we use in this appendix.}

While general, this procedure is not particularly convenient for force-free magnetospheres. Physical magnetosphere models typically contain a current sheet, which appears as a cusp in the flux function. (For example, the split monopole has $\psi=1-\ab{\cos{\theta}}$.) While such functions can be expanded in the series \eqref{eq:SeriesExpansion}, the representation is not intuitive, and the cusp limits the convergence. Furthermore, computing the coefficients $C^\gtrless_\ell$ involves evaluating integrals of the form \eqref{eq:T}, a difficult task in general. To overcome these difficulties, we will use a method appropriate for magnetospheres with equatorial reflection symmetry, where we work only on the northern hemisphere $\theta\in[0,\pi/2]$, with the southern fields to be determined by reflection. This frees us to consider functions $\psi$ that vanish only on one pole, which means the $\ell=0$ harmonic $\Theta_0(\theta)=1-\cos\theta$ is now admissible. The corresponding radial harmonic is $R_0(r)=1$, with the other linearly independent solution divergent at both the horizon and at infinity.

The method takes advantage of the fact that the flat space radial harmonics are pure powers of $r$, $R^<_{\ell}=r^{\ell+1}$, and $R^>_{\ell}=r^{-\ell}$, making Taylor expansion equivalent to mode expansion. We Taylor expand the flat solution about $r=0$ and $r=\infty$ to produce expansions valid for $r\leq b$ and $r\geq b$, respectively. We then promote each expansion to a solution in Schwarzschild by the above procedure. The series converge on either side of a Schwarzschild coordinate radius $r=b_\circ(M,b)$ that we compute explicitly. The overall normalization of the outer solution is fixed by demanding that the flat and Schwarzschild solutions agree asymptotically where both spacetimes are flat, while the overall normalization of the inner expansion is fixed by demanding continuity at $r=b_\circ$. The remainder of this appendix follows this procedure.

\subsection{Angular and Radial Harmonics}
\label{sec:Harmonics}

The separation of Eq.~\eqref{eq:SchwarzschildVacuumEquation} has been discussed previously by Refs.~\cite{Beskin:1992,Ghosh:1999in,Tomimatsu:2000yp}. Our treatment establishes notation for this paper, provides some additional details, and finds the normalization of the radial harmonics required to properly reduce to the canonical normalization in flat spacetime.

The product solutions $R_{\ell}(r)\Theta_{\ell}(\theta)$ of Eq.~\eqref{eq:SchwarzschildVacuumEquation} satisfy 
\begin{align}
	\label{eq:AngularEquation}
	\frac{d}{d\theta}\pa{\frac{1}{\sin{\theta}}\frac{d\Theta_\ell}{d\theta}}
	+\frac{\ell(\ell+1)}{\sin{\theta}}\Theta_\ell&=0\\
	\label{eq:RadialEquation}
	\frac{d}{dr}\br{\pa{1-\frac{2M}{r}}\frac{dR_{\ell}}{dr}}-\frac{\ell(\ell+1)}{r^2}R_\ell&=0, 
\end{align}
where we name the separation constant $\ell(\ell+1)$. Since the angular equation \eqref{eq:AngularEquation} is of Sturm-Liouville form with weight $\csc{\theta}$, we can expect\footnote{The function $\csc{\theta}=1/\sin{\theta}$ is not sufficiently regular for the standard theorems to directly apply.} a discrete, infinite set of eigenvalues whose eigenfunctions are complete and orthogonal with respect to $\csc{\theta}\ed\theta$. We find that the solutions vanishing at both poles are given (in a convenient normalization) by the hypergeometric function,
\begin{align}
	\label{eq:ThetaOdd}
	\Theta_{2k-1}(\theta)&={_2F_1}\!\br{-k,k-\frac{1}{2};\frac{1}{2};\cos^2{\theta}},\\
	\label{eq:ThetaEven}
	\Theta_{2k}(\theta)&={_2F_1}\!\br{-k,k+\frac{1}{2};\frac{3}{2};\cos^2{\theta}}\cos{\theta},
\end{align}
where $k$ (and hence $\ell$) is a positive integer. The $\Theta_\ell$ are proportional to the Gegenbauer polynomials $\mathcal{G}_n^{(m)}(x)$,
\begin{numcases}
	{\Theta_\ell(\theta)=}
		{_2F_1}\!\br{\frac{\ell}{2},-\frac{\ell+1}{2};\frac{1}{2};\cos^2{\theta}}
		=-\frac{\Gamma\!\pa{-\frac{\ell}{2}}\Gamma\!\pa{\frac{\ell+1}{2}}}{2\sqrt{\pi}}
		\sin^2{\theta}\,\mathcal{G}^{(3/2)}_{\ell-1}(\cos{\theta})
		& $\ell$ odd,\nonumber\\
		&\\
		{_2F_1}\!\br{-\frac{\ell}{2},\frac{\ell+1}{2};\frac{3}{2};\cos^2{\theta}}\cos{\theta}
		=-(-1)^{\ell/2}\frac{\sqrt{\pi}\,\Gamma\!\pa{\frac{\ell}{2}}}{4\,\Gamma\!\pa{\frac{\ell+3}{2}}}
		\sin^2{\theta}\,\mathcal{G}^{(3/2)}_{\ell-1}(\cos{\theta})
		&$\ell$ even.\nonumber
\end{numcases}
From the known orthogonality properties of those polynomials, one can directly check that the $\Theta_\ell$ are orthogonal with the expected weight $\csc{\theta}\ed\theta$. Given a sufficiently regular function $T(\theta)$ vanishing at the poles, a series representation can now be computed as 
\begin{align}
	\label{eq:T}
	T(\theta)=\sum_{\ell=1}^\infty T_\ell\Theta_\ell(\theta),\qquad
	T_\ell=\frac{\int_0^\pi T(\theta)\Theta_\ell(\theta)\csc{\theta}\ed\theta}
		{\int_0^\pi\Theta_\ell(\theta)^2\csc{\theta}\ed\theta}.
\end{align}
We will also find it useful to consider the $\ell=0$ solution, which is $\alpha+\beta\cos{\theta}$ for constants $\alpha$ and $\beta$. We can make this function vanish at one pole and hence use it in a single hemisphere. We work with the choice
\begin{align}
	\Theta_0(\theta)=1-\cos{\theta}.
\end{align}
This function is {\it not} orthogonal to the functions $\Theta_{\ell\geq1}$ defined in Eqs.~\eqref{eq:ThetaOdd} and \eqref{eq:ThetaEven}, and we do not regard it as part of that set, which is expected to be complete only for functions vanishing at both poles.

In a convenient normalization, the solutions of the radial equation \eqref{eq:RadialEquation} regular at the horizon and infinity, respectively, are given for $\ell\geq1$ by
\begin{align}
	\label{eq:R<M}
	R^<_\ell(r)&=\frac{r^2}{2}(-2M)^{\ell-1}\frac{\Gamma(\ell+2)^2}{\Gamma(2\ell+1)}\,
		{_2F_1}\!\br{\ell+2,1-\ell;3;\frac{r}{2M}},\\
	R^>_\ell(r)&=-\frac{2}{\sqrt{\pi}}\pa{\frac{r}{4}}^{-\ell}\frac{\Gamma\!\pa{\ell+\frac{3}{2}}}{(\ell+1)\Gamma(\ell)}
		\cu{{_2F_1}\!\br{\ell+2,\ell;1;1-\frac{2M}{r}}\log\pa{1-\frac{2M}{r}}+P_\ell\!\pa{\frac{r}{2M}}},
\end{align}
where the $P_\ell$ are the polynomials recursively defined by
\begin{align}
	P_1(x)&=x^2+\frac{x}{2},\\
	P_2(x)&=4x^4-x^3-\frac{x^2}{6},\\
	P_\ell(x)&=\frac{(2\ell-1)[\ell(\ell-1)(2x-1)-1]xP_{\ell-1}(x)-\ell^2(\ell-2)x^2P_{\ell-2}(x)}{(\ell+1)(\ell-1)^2}.
\end{align}
Note that the leading term in $P_\ell$ is
\begin{align}
	P_\ell(x)=\frac{4^\ell\ell}{\sqrt{\pi}}\frac{\Gamma\!\pa{\ell+\frac{1}{2}}}{\Gamma(\ell+2)}x^{2\ell}
		+\mathcal{O}\!\pa{x^{2\ell-1}}.
\end{align}
The normalization of $R^\gtrless_\ell$ is chosen so that as $r \rightarrow \infty$ fixing $M$ (or equivalently $M\rightarrow 0$ fixing $r$), we reduce to the simple form
\begin{align}
	\label{eq:R0}
	R^<_\ell(r)=r^{\ell+1},\qquad R^>_\ell(r)=r^{-\ell}\qquad\text{as}\qquad
	r\rightarrow\infty\text{ or }M\rightarrow0.
\end{align}
This makes the $M\neq0$ normalization of Eq.~\eqref{eq:R<M} the ``same'' (in the sense discussed in Sec.~\ref{sec:Overview}) as the natural $M=0$ normalization $R^<_\ell(r)=r^{\ell+1}$ and $R^>_\ell(r)=r^{-\ell}$. When $M \neq 0$, $R^<_\ell(r)$ diverges logarithmically on the event horizon, $r=2M$. 

Note that the $M=0$ eigenfunctions in Eq.~\eqref{eq:R0} differ from the usual $r^{\ell}$ and $r^{-(\ell+1)}$ that arise in solutions of the Laplace equation. The reason is that we work with the flux function rather than the magnetostatic scalar potential (see discussion in footnote \ref{FootnoteBis}).

For $\ell=0$, the general solution is $\alpha+\beta[r+2M\log(r-2M)]$ for some constants $\alpha$ and $\beta$. Unlike the $\ell\geq1$ cases, the $\ell=0$ solution is either everywhere regular ($\beta=0$) or divergent at both the horizon and infinity ($\beta\neq 0$).

\subsection{Mode Expansion of the Flat Solution}

Setting $\psi_0=1$ for simplicity, the flat vacuum solution $\psi$, Eq.~\eqref{eq:SolutionOmega0}, is given in spherical coordinates by
\begin{align}
	\label{eq:FlatSpaceVacuumSolution}
	\psi(r,\theta)=1-\sqrt{\frac{1}{2}\br{1-\pa{\frac{r}{b}}^2}+\sqrt{\frac{1}{4}\br{1-\pa{\frac{r}{b}}^2}^2+\pa{\frac{r}{b}}^2\cos^2{\theta}}}.
\end{align}
In principle, one could proceed by integrating $\psi$ against the angular harmonics $\Theta_{\ell}$ from Eq.~\eqref{eq:T} to determine the coefficients $C^\gtrless_{\ell}$ in Eq.~\eqref{eq:SeriesExpansion}, but this approach is intractable. We will take an alternative approach based on the observation that the radial eigenfunctions in flat spacetime are $r^{\ell+1}$ and $r^{-\ell}$ exactly, and therefore correspond to terms in a Taylor expansion around $r=0$ and $r=\infty$, respectively. Thus, by Taylor expanding, we can expect each term to come with an angular function $\Theta_\ell$ and a coefficient that can easily be read off. For example, expanding \eqref{eq:FlatSpaceVacuumSolution} near $r=0$, we find 
\begin{align}
	\label{eq:SmallRadiusFlat}
	\psi(r,\theta)
	=\frac{\sin^2{\theta}}{2}\pa{\frac{r}{b}}^2-\frac{\sin^2{\theta}}{16}\pa{3+5\cos{2\theta}}\pa{\frac{r}{b}}^4
	+\frac{\sin^2{\theta}}{128}\pa{15+28\cos{2\theta}+21\cos{4\theta}}\pa{\frac{r}{b}}^6+\ldots,
\end{align}
each term of which has $\theta$-dependence proportional to an angular function $\Theta_\ell$. The full series is
\begin{align}
	\label{eq:FlatSpaceFluxOrigin}
	\psi(r,\theta)=\frac{1}{2\sqrt{\pi}}\sum^{\infty}_{k=1}
	\frac{\Gamma\!\pa{k-\frac{1}{2}}}{\Gamma(k+1)}\pa{\frac{r}{b}}^{2k}\Theta_{2k-1}(\theta),
	\qquad r\leq b.
\end{align}
We may similarly expand around $r=\infty$ and read off the complete Taylor series as
\begin{align}
	\label{eq:FlatSpaceFluxInfinity}
	\psi(r,\theta)=1-\cos{\theta}-\sum^\infty_{k=1}
	\frac{\Gamma(2k+1)}{\Gamma(k+1)^2}\pa{\frac{b}{2r}}^{2k}\Theta_{2k}(\theta),
	\qquad r\geq b.
\end{align}
Here and hereafter, we work in the northern hemisphere $0\leq\theta\leq\pi/2$. Notice that the near series \eqref{eq:FlatSpaceFluxOrigin} involves only odd multipoles $\ell=2k+1$, while the far series \eqref{eq:FlatSpaceFluxInfinity} involves even multipoles $\ell=2k$. (Note that $1-\cos{\theta}$ is the $\ell=0$ harmonic.) These series do not match term by term and must be summed for comparison at $r=b$.

We have established the radii of convergence of these series by using the ratio test at the special value of $\theta=\pi/2$ and confirming numerically that the radius is $\theta$-independent. This procedure is described in Sec.~\ref{sec:FlatSpaceConvergence} below. Equations~\eqref{eq:FlatSpaceFluxOrigin} and \eqref{eq:FlatSpaceFluxInfinity} provide a dual series representation of the exact flat solution \eqref{eq:FlatSpaceVacuumSolution}.

\subsection{Schwarzschild Solution}

To promote the flat solution to Schwarzschild spacetime, we send $r^{\ell+1}\rightarrow R^<_\ell(r)$ and $r^{-\ell}\rightarrow R^>_\ell(r)$ in Eqs.~\eqref{eq:FlatSpaceFluxOrigin} and \eqref{eq:FlatSpaceFluxInfinity} and also allow for a relative normalization $C$,
\begin{subequations}
	\label{eq:SchwarzschildSolution}
	\begin{numcases}
	{\psi(r,\theta)=}
		\label{eq:HorizonFlux}
		\frac{C}{2\sqrt{\pi}}\sum^{\infty}_{k=1}
		\frac{\Gamma\!\pa{k-\frac{1}{2}}}{\Gamma(k+1)}\pa{\frac{1}{b}}^{2k}R_{2k-1}^<(r)\Theta_{2k-1}(\theta),
		& $r\leq b_\circ$, \\
		\label{eq:InfinityFlux}
		1-\cos{\theta}-\sum_{k=1}^\infty
		\frac{\Gamma(2k+1)}{\Gamma(k+1)^2}\pa{\frac{b}{2}}^{2\ell}R^>_{2k}(r)\Theta_{2k}(\theta),
		& $r\geq b_\circ$.
	\end{numcases}
\end{subequations}
In Sec.~\ref{sec:SchwarzschildConvergence} below, we determine the radius of convergence $b_\circ$ to be
\begin{align}
	\label{eq:DiskRadius}
	b_\circ=b+M+\frac{M^2}{4b}.
\end{align}
This indicates that the disk terminates at the Schwarzschild coordinate $r=b_\circ$. That $b_\circ \neq b$ should not be surprising, since there is no canonical identification between coordinates on different manifolds. (Note, however, that $b_\circ\to b$ as $M\to0$, as required.) Inverting Eq.~\eqref{eq:DiskRadius} gives 
\begin{align}
	\label{eq:b}
	b=\frac{b_\circ-M+\sqrt{b_\circ(b_\circ-2M)}}{2}.
\end{align}
We regard $b_\circ$ as a free parameter in the Schwarzschild solution.

In principle, the normalization $C$ may be determined by summing both series at $r=b_\circ$. This is tractable at the special value of $\theta=\pi/2$ where the angular harmonics simplify. The even harmonics vanish, showing that $\psi(r\geq b)=1$. The odd harmonics are equal to unity, so we have
\begin{align}
	\label{eq:C}
	\frac{1}{C}=\frac{1}{2\sqrt{\pi}}\sum^{\infty}_{k=1}
	\frac{\Gamma\!\pa{k-\frac{1}{2}}}{\Gamma(k+1)}\pa{\frac{1}{b}}^{2k}R_{2k-1}^<(b_\circ).
\end{align}
This series converges excruciatingly slowly, making direct computation with Eq.~\eqref{eq:C} impractical. Its terms only fall off like $k^{-3/2}$, so it is necessary to sum the first $\sim2,600$ terms to get to within 1\% accuracy. In Sec.~\ref{sec:SeriesAcceleration}, we derive an equivalent, more practical representation using series acceleration techniques, Eq.~\eqref{eq:KummerExact}.

To complete the proof that Eq.~\eqref{eq:SchwarzschildSolution} is a smooth solution in Schwarzschild one should in principle check that the series match at all angles $\theta$ using the expression \eqref{eq:C} for $C$ determined at $\theta=\pi/2$. We have verified this numerically in broad brush, but it is difficult to be conclusive because of the slow convergence of at least one of the series at every value of $\theta$. One can expect the series to match because of the unique, natural association between solutions in Schwarzschild and solutions in flat spacetime discussed in Sec.~\ref{sec:Overview}. This association acts over finite neighborhoods and up to overall normalization. Thus, we can expect only a relative normalization when we promote our single flat solution in the two different neighborhoods.

\subsection{Flat Radius of Convergence}
\label{sec:FlatSpaceConvergence}

We now discuss the radius of convergence of the series representations \eqref{eq:FlatSpaceFluxOrigin} and \eqref{eq:FlatSpaceFluxInfinity} of the flat solution. We will need the asymptotic approximation for the hypergeometric function given on p. 77 of Ref.~\cite{Bateman:1953}
\begin{align}
	\label{eq:HypergeometricAsymptotics}
	{_2F_1}\!\br{a+\lambda,b-\lambda;c;\frac{1-z}{2}}
	&\stackrel{\lambda\to\infty}{\sim}
	\frac{\Gamma(1-b+\lambda)\Gamma(c)}{\Gamma\!\pa{\frac{1}{2}}\Gamma(c-b+\lambda)}
	2^{a+b-1}\pa{1-e^{-\xi}}^{-c+1/2}\pa{1+e^{-\xi}}^{c-a-b-1/2}\nonumber\\
	&\quad\times\lambda^{-1/2}\br{e^{(\lambda-b)\xi}+e^{\pm i\pi(c-1/2)}e^{-(\lambda+a)\xi}}
	\br{1+\mathcal{O}\!\pa{\ab{\lambda^{-1}}}},
\end{align}
where the upper or lower sign is taken according to $\Im z\gtrless0$, while $\xi$ is defined by $e^{\xi}\equiv z+\sqrt{z^2-1}$. (Note that $e^{-\xi}=z-\sqrt{z^2-1}$.) For real arguments $z=a+0i$, one considers $z=a\pm i\epsilon$ and lets $\epsilon\rightarrow0$, where the result is independent of the choice of $\pm$.

This formula allows us to find the asymptotic behavior as $k\to\infty$ of \eqref{eq:ThetaOdd} by setting $a=-1/2$, $b=0$, $c=1/2$, $\lambda=k$, and $z=1-2\cos^2{\theta}$, resulting in
\begin{align}
	\label{eq:AngularAsymptotics}
	\Theta_{2k-1}(\theta)\stackrel{k\to\infty}{\sim}
	\frac{\Gamma(k+1)}{\Gamma\!\pa{k+\frac{1}{2}}}\sqrt{\frac{1-e^{2i\theta}}{8k}}
	\br{\pa{-e^{-2i\theta}}^k+\pa{-e^{-2i\theta}}^{-k+1/2}}.
\end{align}
Hence, we may rewrite the sum \eqref{eq:FlatSpaceFluxOrigin} in the asymptotic form
\begin{align}
	\psi(r,\theta)=\sum^{\infty}_{k=1}a_k,\qquad
	a_k\stackrel{k\to\infty}{\sim}
	\frac{\Gamma\!\pa{k-\frac{1}{2}}}{\Gamma\!\pa{k+\frac{1}{2}}}
	\pa{\frac{r}{b}}^{2k}\sqrt{\frac{1-e^{2i\theta}}{32\pi k}}
	\br{\pa{-e^{-2i\theta}}^k+\pa{-e^{-2i\theta}}^{-k+1/2}}.
\end{align}
This expression is real at all $\theta$ but involves a delicate cancelation of complex phases at generic values of $\theta$. The ratio $a_{k+1}/a_k$ in general oscillates wildly on account of these phases, and the ratio test only gives weak bounds on the domain of convergence. However, at the special value of $\theta=\pi/2$,
\begin{align}
	\label{eq:RatioTest}
	L\equiv\lim_{k\to\infty}\ab{\frac{a_{k+1}}{a_k}}
	=\lim_{k\to\infty}\ab{\frac{2k-1}{2k+1}\sqrt{\frac{k}{k+1}}\pa{\frac{r}{b}}^2}
	=\pa{\frac{r}{b}}^2.
\end{align}
Thus, the ratio test shows that the series converges for all $r<b$. The test is inconclusive for $r=b$, but the technical convergence of the series there is unimportant for our purposes, since the value can always be computed by working at $r=b-\epsilon$ and letting $\epsilon\rightarrow0$. For this reason, we write $r\leq b$ as the domain of validity of the series \eqref{eq:FlatSpaceFluxOrigin}. We have not proven that the radius is $r=b$ for all $\theta$, but this is straightforward to verify by numerical experiment. Similarly, one may check numerically that the outer series converges for $r\geq b$.

\subsection{Schwarzschild Radius of Convergence}
\label{sec:SchwarzschildConvergence}

Finding the radii of convergence of the Schwarzschild series \eqref{eq:SchwarzschildSolution} requires the asymptotics of the radial functions \eqref{eq:R<M}, which also follows from Eq.~\eqref{eq:HypergeometricAsymptotics}. For the inner series, we take $a=1$, $b=2$, $c=3$, $\lambda=2k$, and $z=1-r/M$ to find
\begin{align}
	\label{eq:RadialAsymptotics}
	R_{2k-1}^<(r)\stackrel{k \to\infty}{\sim}
	\sqrt{\frac{32}{\pi k}}\frac{\Gamma(2k-1)}{\Gamma(2k+1)}
	\frac{e^{2k \xi}+ie^{-(2k-1)\xi}}{\pa{1-e^\xi}^2\sqrt{1-e^{-2\xi}}},\qquad
	e^\xi=1-\frac{r-\sqrt{r(r-2M)}}{M}.
\end{align}
Hence, we may rewrite the sum as
\begin{align}
	\label{eq:SeriesTerms}
	\psi(r,\theta)=C\sum^{\infty}_{k=1}b_k,\qquad
	b_k\equiv\frac{1}{2\sqrt{\pi}}
	\frac{\Gamma\!\pa{k-\frac{1}{2}}}{\Gamma(k+1)}\pa{\frac{1}{b}}^{2k}R_{2k-1}^<(r)\Theta_{2k-1}(\theta),
\end{align}
and use Eqs.~\eqref{eq:AngularAsymptotics} and \eqref{eq:RadialAsymptotics} to obtain the asymptotic behavior of $b_k$,
\begin{align}
	\label{eq:SeriesAsymptotics}
	b_k\stackrel{k\to\infty}{\sim}\frac{(-1)^k}{2\pi}\pa{\frac{r}{b}}^2
	\frac{\Gamma(2k+1)^2}{\Gamma(4k-1)}\pa{\frac{2M}{b}}^{2(k-1)}
	\sqrt{\frac{1-e^{2i\theta}}{1-e^{-2\xi}}}
	\frac{\pa{e^{2k\xi}+ie^{-(2k-1)\xi}}\pa{e^{-2k i\theta}+ie^{(2k-1)i\theta}}}
	{k^2(2k-1)^2\pa{1-e^\xi}^2}.
\end{align}
As in the asymptotic formula \eqref{eq:AngularAsymptotics} for the angular functions, the ratio test gives only weak bounds for generic values of $\theta$, on account of complex phases. At the special value of $\theta=\pi/2$, however, we have
\begin{align}
	L\equiv\lim_{k\to\infty}\ab{\frac{b_{k+1}}{b_k}}
	=\pa{\frac{M}{2b}}^2\ab{e^{2\xi}}
	=\pa{\frac{M}{2b}}^2\br{1-\frac{r-\sqrt{r(r-2M)}}{M}}^2,
\end{align}
where the last step follows from Eq.~\eqref{eq:RadialAsymptotics}. Demanding $L=1$ gives the formula \eqref{eq:DiskRadius} for $b_\circ$, where $L<1$ for all $r<b_\circ$ so that convergence is assured there. In the same spirit as the discussion below Eq.~\eqref{eq:RatioTest}, we regard $r\leq b_\circ$ as the domain of validity. Similarly, we expect $r\geq b_\circ$ to be the domain of validity of the outer series, which is straightforward to confirm numerically.

\subsection{Series Acceleration for Matching Coefficient}
\label{sec:SeriesAcceleration}

To facilitate numerical evaluation of the normalization $C$ in Eq.~\eqref{eq:HorizonFlux}, we partially resum the series \eqref{eq:C} to produce a more rapidly convergent expression. The series may be written as
\begin{align}
	\frac{1}{C}=\sum^{\infty}_{k=1}c_k,\qquad
	c_k=\frac{1}{2\sqrt{\pi}}
	\frac{\Gamma\!\pa{k-\frac{1}{2}}}{\Gamma(k+1)}\pa{\frac{1}{b}}^{2k}R_{2k-1}^<(b_\circ).
\end{align}
The $c_k$ are equal to the $b_k$ defined in Eq.~\eqref{eq:SeriesTerms} evaluated at $r=b_\circ$ and $\theta=\pi/2$. From the asymptotic behavior \eqref{eq:SeriesAsymptotics}, we have
\begin{align}
	c_k\stackrel{k\to\infty}{\sim}\frac{1}{\pi\sqrt{2\pi}}\frac{b\,b_\circ^2}{\sqrt{2b-M}\pa{2b+M}^{5/2}}
	\frac{2^{6k}\Gamma\!\pa{k-\frac{1}{2}}\Gamma\!\pa{k+\frac{1}{2}}\Gamma(2k-1)}{\sqrt{k}\,\Gamma(4k-1)},
\end{align}\
where we used the fact that at $r=b_\circ$,
\begin{align}
	e^\xi=1-\frac{b_\circ-\sqrt{b_\circ(b_\circ-2M)}}{M}=-\frac{M}{2b},
\end{align}
with $b$ a function of $b_\circ$ as in Eq.~\eqref{eq:b}. This large-$k$ asymptotic behavior is not yet exactly summable, so we appeal to Stirling's approximation $\Gamma(z)\stackrel{\lambda\to\infty}{\sim}\sqrt{2\pi}\,z^{z-1/2}e^{-z}$ in order to further simplify it to
\begin{align}
	c_k\stackrel{k\to\infty}{\sim}d_k,\qquad
	d_k\equiv\frac{4}{\sqrt{\pi}}\frac{b\,b_\circ^2}{\sqrt{2b-M}\pa{2b+M}^{5/2}}\frac{1}{k^{3/2}}.
\end{align} 
We now recognize the weak $k^{-3/2}$ falloff as the reason behind the slow convergence of the series. But the sum of $k^{-n}$ is the Riemann zeta function $\zeta(n)$, so we have
\begin{align}
	\sum_{k=1}^\infty d_k
	=\frac{4}{\sqrt{\pi}}\frac{b\,b_\circ^2}{\sqrt{2b-M}\pa{2b+M}^{5/2}}\zeta\!\pa{\frac{3}{2}}.
\end{align}
Thus, we may instead write
\begin{align}
	\label{eq:KummerExact}
	\frac{1}{C}=\frac{4}{\sqrt{\pi}}\frac{b \ \! b_\circ^2}{\sqrt{2b-M}\pa{2b+M}^{5/2}}\zeta\!\pa{\frac{3}{2}}
	+\sum_{k=1}^\infty\pa{b_k-c_k}.
\end{align}
The series of correction terms converges quite quickly. If we keep only the first term $k=1$, the estimate of $C$ is
\begin{align}
	\label{eq:KummerFirstTerm}
	\frac{1}{C}=\frac{b_\circ^2}{2b^2}
	+\frac{4}{\sqrt{\pi}}\frac{b\,b_\circ^2}{\sqrt{2b-M}\pa{2b+M}^{5/2}}\br{\zeta\!\pa{\frac{3}{2}}-1}.
\end{align}
This truncated expression is generally accurate to within about 5\%. For example, if the disk extends to the ISCO, $b_\circ=r_{\text{ISCO}}=6M$, we get an exact value of $C^{-1}=1.3603...$ from Eq.~\eqref{eq:KummerExact} and an approximate value of $C^{-1}\approx1.289$ from Eq.~\eqref{eq:KummerFirstTerm}. For $b\gg M$, we know that $C=1$ exactly, and Eq.~\eqref{eq:KummerFirstTerm} gives $C^{-1}\approx.955$.\hfill\includegraphics[scale=0.03]{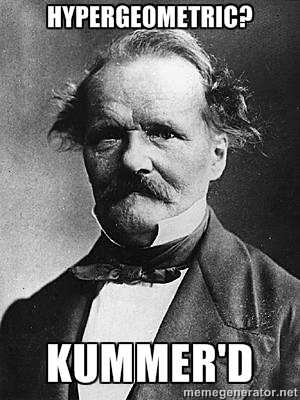}

\bibliography{AGN}
\bibliographystyle{utphys}

\end{document}